\documentclass[fleqn,usenatbib]{mnras}

\usepackage{newtxtext,newtxmath}
\usepackage[T1]{fontenc}

\DeclareRobustCommand{\VAN}[3]{#2}
\let\VANthebibliography\thebibliography
\def\thebibliography{\DeclareRobustCommand{\VAN}[3]{##3}\VANthebibliography}

\usepackage{graphicx}	% Including figure files
\usepackage{amsmath}	% Advanced maths commands
\usepackage[none]{hyphenat}
\usepackage{longtable}
\usepackage{booktabs}
\usepackage{blindtext}
\usepackage{natbib}
\usepackage{dblfloatfix}
\title[Recovering halo masses from clustering]{Recovering the characteristic halo mass of rare populations from clustering: a benchmark using the Millennium Simulation}

\author[A. Banerjee et al.]{
Amrita Banerjee,$^{1,2}$\thanks{E-mail: amrita.banerjee.physics@gmail.com}
Darren Croton,$^{1,2}$
\\
$^{1}$Centre for Astrophysics and Supercomputing, Swinburne University of Technology, PO Box 218, Hawthorn VIC 3122, Australia\\
$^{2}$ARC Centre of Excellence for Dark Matter Particle Physics (CDM) , Melbourne 3000, Australia\\}

\date{Accepted XXX. Received YYY; in original form ZZZ}

\pubyear{\the\year{}}

\begin{document}
\label{firstpage}
\pagerange{\pageref{firstpage}--\pageref{lastpage}}
\maketitle

\begin{abstract}
We measure how accurately the characteristic halo mass of a rare, massive population can be recovered from clustering alone, using dark matter haloes in the Millennium simulation, where the true masses are known. Massive samples of 25 to 10000 haloes are cross-correlated against field reference samples of 5000 to 50000 haloes at six redshifts over $0 \leq z \leq 5.29$. The bias is measured at $r = 8\,h^{-1}\,\mathrm{Mpc}$ and inverted to return a characteristic mass, which we compare with the mean logarithmic halo mass measured directly from the simulation. Across 192 configurations the characteristic mass is recovered with a median absolute offset of $0.103$ dex. We show that (i) accuracy depends only weakly on the number of rare objects while the uncertainty falls from $\pm 0.71$ dex at 25 objects to $\pm 0.08$ dex at 10000; (ii) enlarging the smallest samples gives the greatest return, with a doubling from 25 to 50 objects reducing the uncertainty by about a third, against a tenth from 5000 to 10000; and (iii) the reference population carries an error of its own, which propagates coherently into every cross-correlation measured against it and is not reduced by observing more rare objects. Working with haloes rather than galaxies isolates the clustering measurement from the galaxy--halo connection, providing a benchmark before population-specific complications are introduced. For the best configurations, the uncertainty contributed by the calibration of the bias--mass relation, at $\sim 0.07$ dex, already matches that of the measurement itself.
\end{abstract}


%%%%%%%%%%%%%%%%%%%%%%%%%%%%%%%%%%%%%%%%%%%%%%%%%%

%%%%%%%%%%%%%%%%% BODY OF PAPER %%%%%%%%%%%%%%%%%%

\section{Introduction}
\label{sec:introduction}

In the $\Lambda$CDM framework galaxies form within dark matter haloes that grow hierarchically from primordial density fluctuations. The most massive haloes collapse from the rarest peaks of the initial density field, and because those peaks are more strongly clustered than the matter as a whole, the bias of a halo population rises steeply with mass and, at fixed mass, with redshift \citep{Kaiser1984,Mo1996}. Halo mass sets the depth of the potential well in which a galaxy forms, and thereby influences the accretion, cooling and retention of its baryons. It is the natural quantity against which to test a model of galaxy formation. It is also not directly observable.

Several routes exist for estimating halo mass. Weak gravitational lensing measures the projected mass around a population directly, but typical galaxy-scale haloes require stacking many objects to reach useful precision \citep{Mandelbaum2006}. Satellite kinematics require spectroscopy of companions \citep{More2011}. Abundance matching infers halo mass by matching galaxy and halo abundances under an assumed correspondence between the two, and so cannot test that relation independently \citep{Behroozi2010,2018ARA&A..56..435W}. For rare populations at high redshift none of these are straightforward, and clustering is frequently the most practical route.

Clustering estimates halo mass through bias. The two-point correlation function of a population is compared with that of the underlying matter, the square root of their ratio gives the bias, and a bias--mass relation calibrated on simulations is inverted to return a characteristic mass \citep{Tinker2010,Desjacques2018}. Observationally, the method requires only positions and a reference against which to measure them, which makes it practical in regimes where other methods are difficult. For sparse populations, the autocorrelation is poorly determined, and cross-correlating against a denser reference population retains signal where the autocorrelation does not.

Populations of exactly this kind are now being found at increasingly high redshift. JWST has revealed massive quiescent galaxies well beyond the redshifts previously accessible \citep{Weibel2025}, luminous quasars are known to $z \approx 7.6$ \citep{Wang2021}, and little red dots are being identified across several deep fields \citep{Matthee2024}, with the first clustering measurements now appearing \citep{Carranza25}. Clustering-based halo masses are already being derived for rare high-redshift populations: \citet{Arita2023} obtain a characteristic mass for $z \approx 6$ quasars from 107 objects, \cite{Eilers2024} measure quasar–galaxy clustering around luminous quasars at $z \geq 6$, while \citet{Schindler2026} constrain quasar--galaxy clustering at $z \simeq 7.3$ from eight companion galaxies across two fields. These measurements sit at the small end of the range of sample sizes considered here, and in places below it.

How well the recovery performs for such small samples has been forecast for specific surveys. \citet{Endsley2020}, for example, predict halo mass precisions of 0.2--0.3 dex from JWST Cycle 1 angular clustering at $4 \leq z \leq 10$. The difficulty is partly structural: in observational work the true halo masses are inaccessible, so a measurement uncertainty can be estimated while the offset between the inferred characteristic mass and the population's true mean cannot be measured directly. Mock-based forecasts provide an important test, but necessarily adopt a model connecting galaxies to their host haloes. What an observer actually faces --- how the accuracy and precision depend on sample size, what a further allocation of telescope time would buy, and at what point the calibration of the bias--mass relation rather than the measurement becomes the limiting term --- is therefore the question we address here.

A simulation removes that obstacle because the true masses are known. In this work we measure how accurately the mean logarithmic halo mass of a rare, massive population is recovered from clustering, using dark matter haloes in the Millennium simulation \citep{Springel2005}. Massive samples of 25 to 10000 haloes are cross-correlated against reference samples of 5000 to 50000 haloes at six redshifts between $z=0$ and $z=5.29$, the resulting bias is inverted through the \citet{Tinker2010} relation, and the recovered mass is compared with the mass measured directly from the simulation. The 192 configurations that result span the sparse-sample regime relevant to current and forthcoming surveys.

Forecasts of this kind have previously been made for particular surveys and particular selections (e.g. \citealt{Endsley2020}), while existing measurements are tied to the samples and fields actually observed (\citealt{Arita2023,Schindler2026}); both necessarily carry the assumptions those choices require. Here we instead vary the size of the rare population, the size of the reference population and the redshift together, so that the behaviour of the method can be mapped across the parameter space. This separates the accuracy of the recovery from its precision, and both from the error introduced by the reference population, which cannot be reduced by observing more rare objects.

Working with haloes rather than galaxies is deliberate. A galaxy analysis must map observed properties onto halo mass, introducing scatter and model dependence in the galaxy--halo connection. Removing these isolates the error of the clustering measurement from that of the galaxy--halo mapping, and establishes what the method can achieve before any population-specific complication is introduced. The results are therefore a benchmark rather than a prediction for any particular population.

This paper is organised as follows: Section~\ref{sec:simulation} describes the simulation and the construction of the halo samples. Section~\ref{sec:methods} sets out the correlation function, bias and mass estimators. Section~\ref{sec:results} presents the recovered masses, their accuracy and their precision across the full set of configurations. Section~\ref{sec:discussion} discusses what these results imply for the design of surveys targeting rare objects, together with the limitations of the approach. Section~\ref{sec:summary} summarises and outlines future work.

\section{N-body DM-only Simulation}
\label{sec:simulation}

\subsection{Millennium simulation}
\label{sec:data}
We build our analysis on the Millennium simulation \citep{Springel2005}, a dark-matter-only cosmological $N$-body simulation of the concordance $\Lambda$CDM cosmogony and a long-standing benchmark for studies of hierarchical structure formation. The simulation follows $2160^3 \simeq 10^{10}$ dark matter particles within a periodic cube of comoving side length $500\,h^{-1}\,\mathrm{Mpc}$, corresponding to a particle mass of $8.6 \times 10^8\,h^{-1}\,\mathrm{M_\odot}$, and softens the gravitational force on a comoving Plummer-equivalent scale of $5\,h^{-1}\,\mathrm{kpc}$, which we take as the spatial resolution of the calculation. At this mass resolution, haloes at our field-sample limit of $10^{11}\,h^{-1}\,\mathrm{M_\odot}$ are resolved with at least $\sim100$ particles, while the massive haloes whose clustering we measure contain thousands to millions, so that halo masses and positions, the only properties our method requires, are robustly determined. The adopted cosmological parameters are consistent with a combined analysis of the 2dFGRS \citep{Colless2001} and first-year \textit{WMAP} data \citep{Spergel2003}: $\Omega_{\mathrm{m}} = 0.25$, $\Omega_{\mathrm{b}} = 0.045$, $\Omega_{\Lambda} = 0.75$, $h = 0.73$, $n_{\mathrm{s}} = 1$ and $\sigma_8 = 0.9$.

\begin{figure*}
    \centering
    \includegraphics[width=\textwidth]{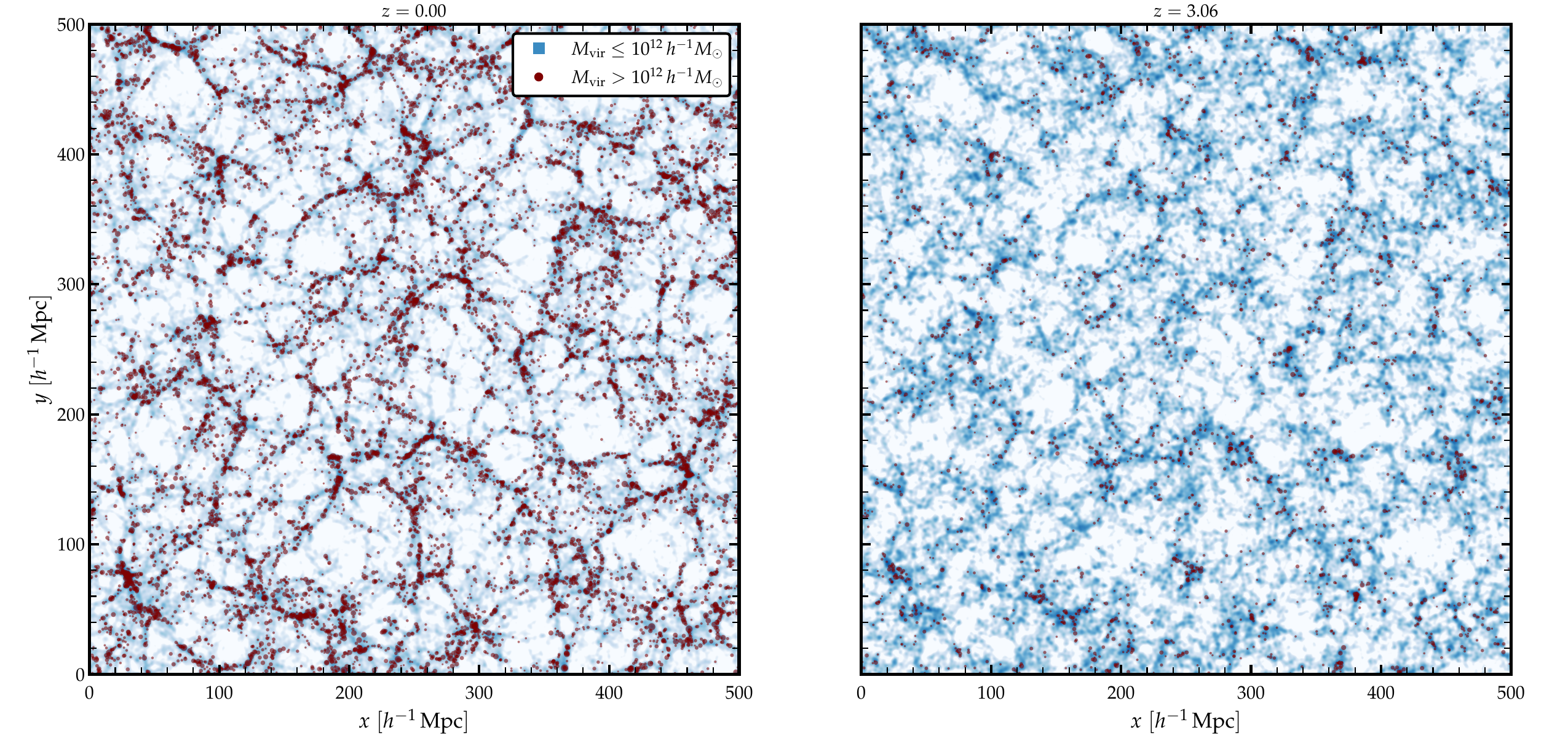}
    \caption{Spatial distribution of dark matter haloes in a slab of thickness $15\,h^{-1}\,\mathrm{Mpc}$ cut from the full Millennium simulation volume, shown at $z = 0$ (left) and $z = 3.06$ (right). Blue points mark haloes with $M_{\mathrm{vir}} \leq 10^{12}\,h^{-1}\,\mathrm{M_\odot}$, tracing the field population; red points mark haloes with $M_{\mathrm{vir}} > 10^{12}\,h^{-1}\,\mathrm{M_\odot}$, where $M_{\mathrm{vir}}$ denotes the spherical-overdensity mass $M_{200\mathrm{c}}$ defined in Section~\ref{sec:methods}. The massive haloes occupy the densest nodes and filaments of the cosmic web at both epochs, visibly clustering more strongly than the general population, as quantified by the bias measurements in Section~\ref{sec:bias}.}
    \label{fig:spatial_distribution}
\end{figure*}

Initial conditions are generated at redshift $z = 127$ by imposing a Gaussian random field, with a linear power spectrum computed by the Boltzmann code \texttt{CMBFAST} \citep{Seljak1996}, on a uniform glass-like particle load \citep{White1996}. The simulation is then evolved to the present day with a memory-optimised TreePM version of the \texttt{GADGET-2} code \citep{Springel2005Gadget}. A total of 64 output snapshots are saved, spaced approximately logarithmically in scale factor, giving a temporal resolution of $\sim300\,\mathrm{Myr}$ by redshift zero and allowing halo assembly histories to be traced in detail.

Dark matter haloes are identified using the standard friends-of-friends (FoF) algorithm with a linking length of 0.2 times the mean interparticle separation \citep{Davis1985}, and gravitationally bound substructures within these haloes are identified using the \texttt{SUBFIND} algorithm \citep{Springel2001}. Halo catalogues and merger trees are constructed using a combination of \texttt{SUBFIND} and the \texttt{L-HaloTree} algorithm \citep{Springel2005}, which links each subhalo to the descendant containing most of its most-bound particles, so that every subhalo retains an unbroken link to its progenitor. These catalogues supply the halo samples analysed throughout this work.

The Millennium simulation offers several practical advantages for this work. Its legacy in the structure formation literature is substantial. Its data products are publicly accessible through the Millennium database \citep{Lemson2006}, their behaviour is thoroughly characterised, and results obtained from the simulation can be placed directly in the context of two decades of published work. Its volume is equally well suited to our purpose. The $500\,h^{-1}\,\mathrm{Mpc}$ box is large enough to contain statistical samples of the rare, massive haloes that dominate the clustering signal, yet comparable to the comoving volumes probed by deep extragalactic surveys, which target fields of limited extent in right ascension and declination rather than the full sky. The simulation therefore provides a realistic parent volume from which survey-like samples can be drawn and against which clustering measurements of comparable volume can be interpreted.

%--------------------------------------------
%--------------------------------------------

\subsection{Halo samples}
\label{sec:halo_samples}

Our analysis uses six snapshots, at $z = 0$, 1.08, 2.07, 3.06, 4.18 and 5.29, spanning the interval over which deep surveys detect rare massive objects. Two families of halo sample are drawn from each. The field sample provides a reference population against which clustering is measured, and the massive sample stands for the rare objects a survey would target.

\begin{figure*}
    \centering
    \includegraphics[width=\textwidth]{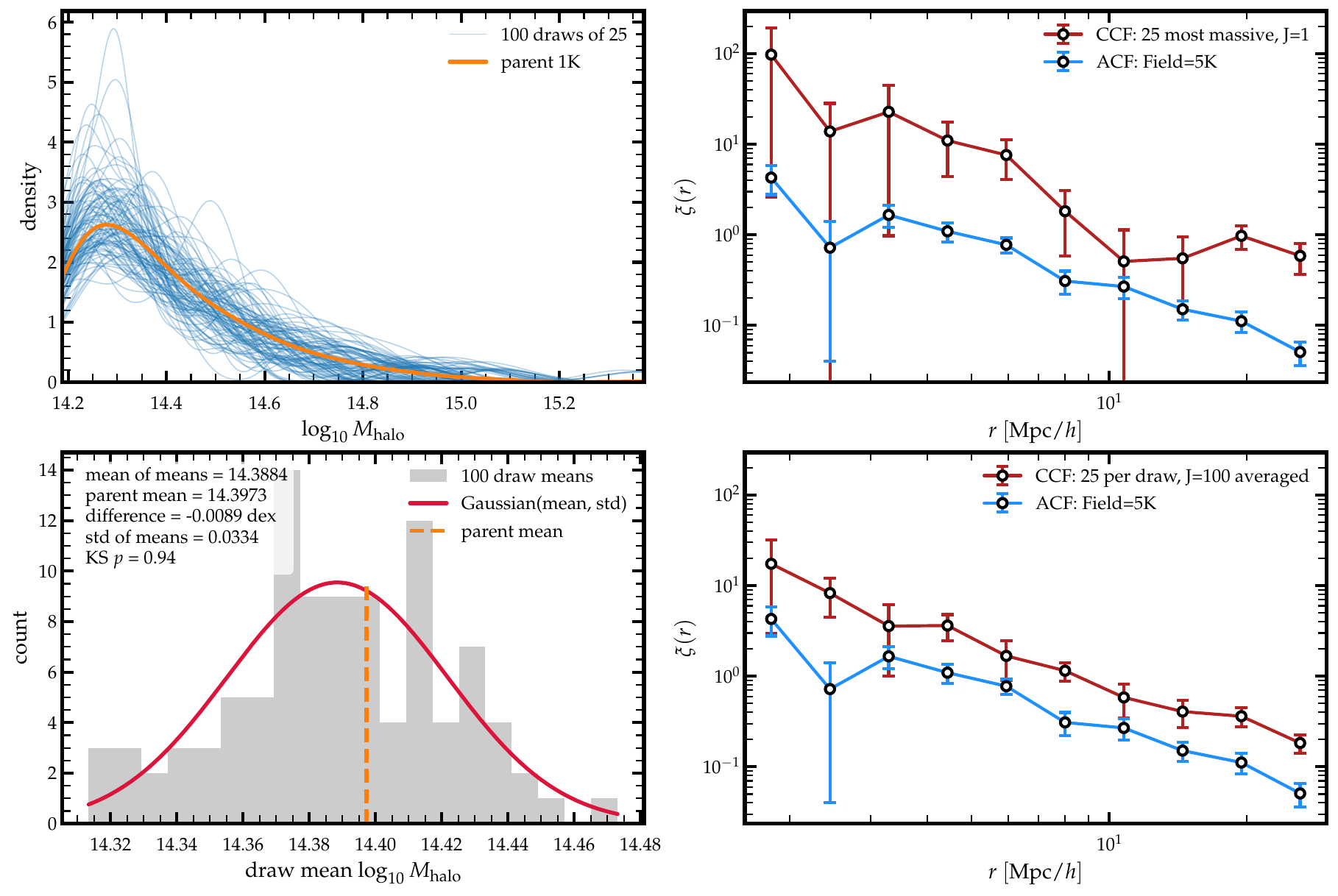}
    \caption{Validation of the $J=100$ sampling procedure at $z=0$. Top left: halo-mass distributions of 100 independent random draws of 25 haloes from the 1000 most massive (thin blue), with the parent distribution overlaid (orange). Individual draws fluctuate around and collectively reproduce the skewed parent distribution. Bottom left: distribution of the mean halo masses of the 100 draws, compared with a Gaussian of the same mean and standard deviation; the parent mean is marked by the dashed line. The mean of the draw means differs from the parent mean by 0.009 dex, with a standard deviation of 0.033 dex. Right: cross-correlation functions for a single deterministic selection of the 25 most massive haloes (top) and the average of 100 random draws of 25 from the top 1000 (bottom), each measured against the 5000-halo field; the field auto-correlation is shown for reference. The two selections represent different populations, so their amplitudes should not be compared directly. Averaging over 100 draws reduces the fractional uncertainty from order unity or greater in some bins to a few tens of per cent across most of the measured range.}
    \label{fig:halo_distribution}
\end{figure*}

Field samples are drawn at random from all haloes above $10^{11}\,h^{-1}\,\mathrm{M_\odot}$, in sizes of 50000, 20000, 10000 and 5000 haloes, with no upper mass limit imposed. The drawn samples reach $\log M = 14.9$ at $z = 0$ and 12.7 at $z = 5.29$; the field is dominated by low-mass haloes without being restricted to them. One field sample is drawn per size and redshift, and every massive sample measured against a given field shares the same reference. Because the field is drawn from the full halo population without an upper limit, a small fraction of the massive haloes appear in it as well; objects common to both samples pair only at zero separation, which lies below the smallest bin used, so they do not contribute to the measured correlation functions. Figure~\ref{fig:spatial_distribution} shows both populations in a slab of the simulation volume at $z = 0$ and $z = 3.06$.

Massive samples are constructed in two ways. For sizes of 10000, 5000, 2000 and 1000 haloes the sample is the $M$ most massive haloes in the box, a single deterministic selection with no sampling freedom; these are denoted $J = 1$. At $z = 0$ they begin at $\log M = 13.51$, 13.73, 14.00 and 14.19 respectively, and all share the same maximum of 15.37. For sizes of 500, 100, 50 and 25 the sample is drawn at random from the 1000 most massive haloes and the procedure repeated 100 times; these are denoted $J = 100$. The correlation functions of the 100 draws are averaged before any bias or mass is computed, so each $J = 100$ configuration yields a single measurement.

The two schemes answer different questions. A deterministic top-$M$ selection asks how well clustering recovers the mass of a complete mass-ranked sample. The $J = 100$ samples ask what happens when only part of that sample is available: a draw of 500, 100, 50 or 25 from the top 1000 corresponds to recovering 50, 10, 5 or 2.5 per cent of it. Averaging over draws separates the sampling from the measurement, at the cost of describing an ensemble rather than any single observation, a distinction returned to in Section~\ref{sec:sample_size_requirements}.

Draw seeds depend on the redshift and the draw size but not on the field size, so the same set of 100 daughter samples is measured against all four field samples. Comparisons across field size are therefore paired, and any difference between them originates in the field rather than in the massive sample.

Figure~\ref{fig:halo_distribution} validates this sampling procedure in its most demanding case, 25 haloes drawn from the top 1000 at $z = 0$. The individual draw distributions bracket the parent pool. The mean of the 100 draw means lies 0.009 dex from the parent mean with a standard deviation of 0.033 dex, and a Kolmogorov--Smirnov test finds the distribution consistent with a Gaussian of the same width ($p = 0.94$). The right-hand panels show what the averaging achieves. The two samples are not quite the same population, the 25 most massive haloes being somewhat more massive than a random 25 drawn from the top 1000, so the comparison is between their uncertainties rather than their amplitudes. A single sample of the 25 most massive haloes yields a cross-correlation with errors spanning an order of magnitude, while the average of 100 draws of 25 is measured to a few tens of per cent across the same range. The smallest samples are analysed as $J = 100$ throughout for this reason: averaging substantially reduces the sampling noise present in individual samples of this size.

Combining eight massive sizes, four field sizes and six redshifts gives 192 configurations, each analysed independently. 

\section{Estimating the average halo mass from clustering analysis}
\label{sec:methods}

\subsection{Two-point Correlation Function}
\label{sec:2pcf}
The two-point correlation function (2PCF) is a standard statistic for quantifying the clustering of a tracer population. Here we compute the spatial 2PCF, which measures the excess probability over random of finding a halo pair at comoving separation $r$ \citep{Peebles1974}. We compute the 2PCF using the Landy--Szalay estimator \citep{Landy1993}, which provides a minimum-variance estimate of $\xi(r)$ for the sample sizes and geometries considered here:

\begin{equation}
\xi(r) =
\frac{DD(r)-2DR(r)+RR(r)}
{RR(r)},
\label{eq:ls_acf}
\end{equation}
where $DD(r)$, $DR(r)$ and $RR(r)$ are the normalised data--data, data--random and random--random pair counts, respectively. Uncertainties are estimated via jackknife resampling of the simulation box volume into 27 spatial subvolumes, which captures both shot noise and sample variance across the scales of interest \citep{2009MNRAS.396...19N}.

For rare populations such as the most massive haloes at high redshift, however, the auto-correlation function becomes noise-dominated. Even for samples drawn from deep extragalactic surveys such as those now conducted with \textit{JWST}, the number densities of the rare massive systems they target fall far below those of the general population, and the resulting shot noise renders the auto-correlation signal unreliable. We therefore measure the cross-correlation function (CCF) between the most massive haloes (MMH) sample and a reference field halo (FH) population, which preserves the clustering signal of the rare population while avoiding or at least minimising the shot-noise penalty of the auto-correlation \citep{DavisPeebles1983,Szapudi1998,Coil2009}:

\begin{equation}
\xi_{AB}(r) = \frac{D_A D_B(r)-D_A R_B(r)-D_B R_A(r)+R_A R_B(r)}{R_A R_B(r)},
\label{eq:ls_ccf}
\end{equation}
where $A$ denotes the field sample and $B$ the MMH population. Two independent random catalogues are generated, one for each sample, mirroring standard observational practice. Pair counts are computed using the \texttt{Corrfunc} code \citep{Sinha2020} and independently verified with our own pair-counting implementation, which yields identical results.

\subsection{Bias}
\label{sec:bias}
The halo bias quantifies how strongly a halo population traces the underlying dark matter distribution. On large scales, where density fluctuations remain in the linear regime, the bias is expected to be approximately scale-independent, with values greater than unity indicating that a population clusters more strongly than the dark matter \citep{Kaiser1984,Desjacques2018}. In hierarchical structure formation, the clustering strength of dark matter haloes increases with halo mass, reflecting the fact that the most massive haloes correspond to rare peaks in the initial density field \citep{Kaiser1984,Mo1996,Sheth1999,Tinker2010}; the MMH samples studied here should therefore exhibit a correspondingly large clustering bias. We evaluate the bias at a comoving scale of $8\,h^{-1}\,\mathrm{Mpc}$, a scale close to the transition between the linear and non-linear regimes and corresponding to the conventional normalisation scale of the matter power spectrum, $\sigma_8$ \citep{Peebles1980}. The bias parameter is defined as

\begin{equation}
b = \sqrt{\frac{\xi(8,z)}{\xi_{\mathrm{DM}}(8,z)}}.
\label{eq:bias_acf}
\end{equation}
Here, $\xi(r)$ is the halo correlation function measured from the simulation and $\xi_{\mathrm{DM}}(r)$ is the dark matter correlation function. The latter is computed using the \texttt{halomod} code \citep{Murray2013,Murray2021} with the halo-model formalism \citep{Cooray2002}, adopting the halo bias model of \citet{Tinker2010}, the CAMB transfer function \citep{Lewis2000,Lewis2011}, and the linear growth-factor formulation of \citet{Carroll1992}.

In the noise-dominated regime identified above, we instead estimate the MMH bias from the cross-correlation with the field population, defining

\begin{equation}
b_{\mathrm{CCF}}(r) = \sqrt{\frac{\xi_{\mathrm{MMH,FH}}(r)}{\xi_{\mathrm{DM}}(r)}},
\label{eq:bias_ccf_1}
\end{equation}
evaluated at the same reference scale. In the linear regime, the cross-correlation amplitude scales as the geometric mean of the biases of the two samples \citep{Mo1996,Mountrichas2009},

\begin{equation}
b_{\mathrm{CCF}}\approx\sqrt{b_{\mathrm{MMH}}\,b_{\mathrm{FH}}},
\label{eq:bias_ccf_2}
\end{equation}
so that the MMH bias is recovered as

\begin{equation}
b_{\mathrm{MMH}} = \frac{b_{\mathrm{CCF}}^2}{b_{\mathrm{FH}}},
\label{eq:bias_massive}
\end{equation}
where $b_{\mathrm{FH}}$ is measured from the field auto-correlation function, which remains signal-dominated owing to the high number density of the field sample. This extends the analysis to regimes where the low number density of the MMH sample would otherwise prevent a reliable direct determination.

\subsection{Characteristic halo mass determination from clustering analysis}
\label{sec:halo_recovery_method}
The dependence of bias on peak significance provides the route from the measured clustering back to a characteristic halo mass: where the bias rises steeply with mass, as it does for the rare peaks considered here, a measured bias selects a narrow range of halo mass at each redshift and the relation can be inverted. The steepness varies along the relation, however, and this variation propagates into the uncertainty of the inferred mass (see Section~\ref{sec:sample_size}).

The significance of the density peak that collapses into a halo of mass $M$ at redshift $z$ is quantified by the peak height,

\begin{equation}
\nu = \frac{\delta_{\mathrm{c}}}{\sigma(M,z)},
\label{eq:nu}
\end{equation}
where $\delta_{\mathrm{c}} = 1.686$ is the linear collapse threshold \citep{Press1974} and $\sigma(M,z)$ is the rms variance of the density field smoothed on the scale corresponding to mass $M$,

\begin{equation}
\sigma^{2}(M,z) = \frac{1}{2\pi^{2}}\int
k^{2}P(k,z)\left|W(kR)\right|^{2}\,\mathrm{d}k,
\label{eq:sigma_peak}
\end{equation}
with $P(k,z)$ the linear matter power spectrum evaluated for the Millennium simulation cosmology, and $W(kR)$ the Fourier-space top-hat filter,

\begin{equation}
W(kR) = \frac{3}{(kR)^{3}}\left[\sin(kR)-kR\cos(kR)
\right],
\label{eq:window_func}
\end{equation}
where the smoothing radius $R$ and halo mass are related by

\begin{equation}
M =
\frac{4\pi}{3}\rho_{\mathrm{m}}R^{3},
\label{eq:mass_from_sigma}
\end{equation}
with $\rho_{\mathrm{m}}$ the mean matter density evaluated for the same cosmology. Large values of $\nu$ correspond to massive haloes that form in the highest-density peaks and carry the largest bias.

To map $\nu$ to a large-scale bias, we adopt the fitting function of \citet{Tinker2010}, calibrated against $N$-body simulations and parameterised continuously in overdensity and redshift,

\begin{equation}
b(\nu) = 1-A\frac{\nu^{a}}{\nu^{a}+\delta_{\mathrm{c}}^{a}}+B\nu^{b}+C\nu^{c},
\label{eq:bias_fitting_func}
\end{equation}
where $(A,a,B,b,C,c)$ are empirical fitting coefficients, tabulated as functions of overdensity in Table~2 of \citet{Tinker2010}. Halo masses are defined as spherical-overdensity masses within a radius enclosing 200 times the critical density, $M_{200\mathrm{c}}$, consistent with the mass definition adopted for the Millennium halo catalogue. We evaluate the \citet{Tinker2010} bias relation for this mass definition, computing $\sigma(M,z)$, $\nu$, and $b(M,z)$ with the \texttt{Colossus} package \citep{Diemer2018}, yielding a direct mapping between halo mass and large-scale bias. The bias of each MMH sample, measured from the clustering analysis described above, is then used to invert this relation and determine the bias-inferred characteristic halo mass of the population.

\section{Results}
\label{sec:results}
Combining eight massive-sample sizes, four field sizes and six redshifts gives 192 configurations, each analysed independently; Table~\ref{tab:1acf} and a representative subset of the fitted correlation parameters, biases and recovered masses are given in Tables~2--5, with the complete set in the supplementary material. To keep the figures readable, we highlight three configurations spanning the range of measurement quality. The best case uses 10000 massive haloes against a 50000-halo field, the typical case 500 against 20000, and the poor case 50 against 10000. These labels refer only to position within the grid; all trends discussed below are drawn from the full set of 192 configurations.

\subsection{Clustering and Bias}
\label{sec:clustering_bias}

\begin{figure*}
    \centering
    \includegraphics[width=\textwidth]{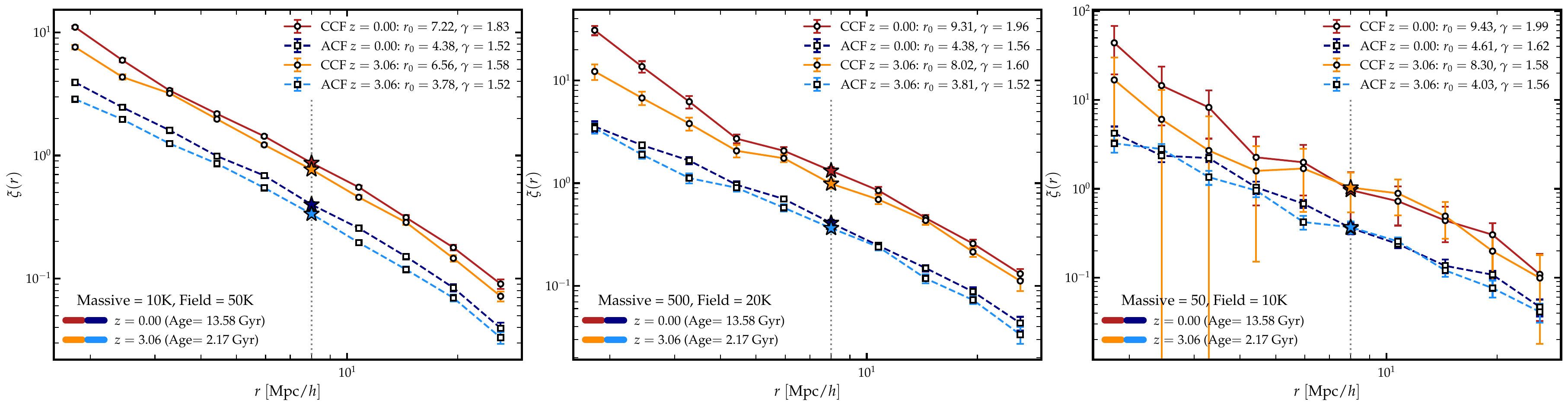}
    \caption{Cross-correlation functions between the massive samples and the field (circles, solid) and field auto-correlations (squares, dashed), for three sample-size configurations---best, typical and poor: 10000 massive haloes against a 50000-halo field, 500 against 20000, and 50 against 10000. At $z=0$ the cross-correlation is drawn in dark red and the auto-correlation in navy; at $z=3.06$ they are drawn in orange and light blue, as identified in the lower legend. Correlation lengths and slopes from a power-law fit are given in the upper legend. The dotted line marks $r=8\,h^{-1}\,\mathrm{Mpc}$, where bias is evaluated. Cross-correlation uncertainties are jackknife estimates for J=1, and the larger of the jackknife and draw-to-draw estimates for J=100; field auto-correlation uncertainties are jackknife. The panels are scaled independently in $\xi(r)$. The overall correlation-function shape is retained across the configurations, while the uncertainties grow markedly from left to right as the massive sample shrinks.}
    \label{fig:cf}
\end{figure*}

The cross-correlation between the massive haloes and the field follows a power law in every configuration and at all redshifts. Only its precision changes. Figure~\ref{fig:cf} shows three representative cases at $z = 0$ and $z = 3.06$: 10000 massive haloes against a 50000-halo field, 500 against 20000, and 50 against 10000, each with the field auto-correlation measured against the same field sample. Table~\ref{tab:2ccf_j1} and \ref{tab:3ccf_j100} list the fitted parameters for the cross-correlations, and Table~\ref{tab:1acf} those for the field auto-correlations.

Smaller massive samples cluster more strongly. At $z = 0$ the cross-correlation lengths are $r_0 = 7.22$, $9.31$ and $9.43\,h^{-1}\,\mathrm{Mpc}$, with slopes $\gamma = 1.83$, $1.96$ and $1.99$. The 500- and 50-halo samples are drawn from the 1000 most massive haloes, while the 10000-halo sample reaches well below that limit. Mass selection sets the ordering, not measurement quality. The same pattern holds at $z = 3.06$, where $r_0 = 6.56$, $8.02$ and $8.30\,h^{-1}\,\mathrm{Mpc}$. The field auto-correlations are far weaker throughout: $r_0 = 4.38$, $4.38$ and $4.61\,h^{-1}\,\mathrm{Mpc}$, with $\gamma = 1.52$, $1.56$ and $1.62$.

Clustering amplitudes fall only slightly between the two epochs. The field auto-correlation at $r = 8\,h^{-1}\,\mathrm{Mpc}$ drops by 16 per cent from $z = 0$ to $z = 3.06$, and the cross-correlation by about 12 per cent. Over the same interval the field bias rises from $0.80$ to $2.23$, which places the mass correlation at roughly 11 per cent of its $z = 0$ amplitude, smaller by a factor of nine. Massive haloes retain a similar clustering amplitude while the underlying matter clustering declines strongly, and the bias measures that difference rather than either quantity alone.

The field auto-correlation barely changes between panels at fixed redshift. The same population is being sampled in each. Its uncertainty at $r = 8\,h^{-1}\,\mathrm{Mpc}$ does change, from 4 per cent for the 50000-halo field to 7 per cent at 20000 and 15 per cent at 10000. This uncertainty enters every configuration equally, and no increase in the massive sample removes it.

\begin{figure*}
    \centering
    \includegraphics[width=\textwidth]{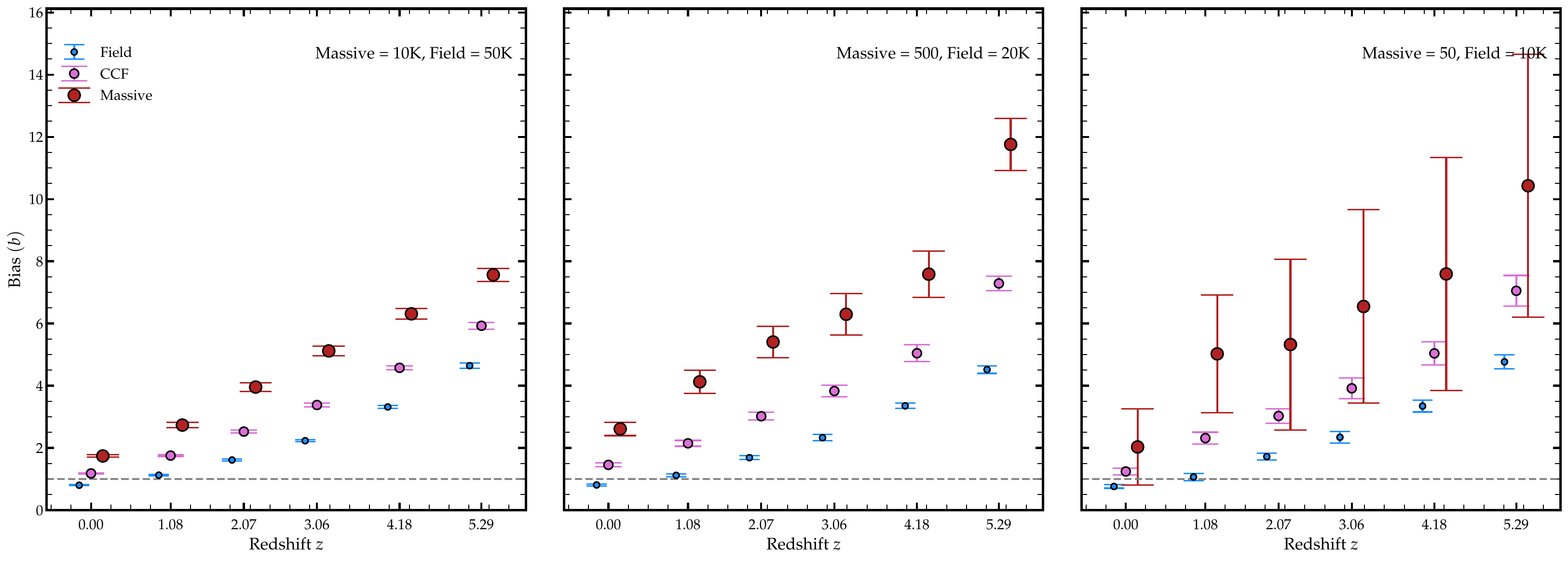}
    \caption{Bias as a function of redshift for the three standard configurations -- best, typical, and poor. Each panel shows the field bias (blue), cross-correlation bias (violet) and massive-halo bias (red), with increasing marker size to distinguish the three quantities where they overlap. Points at each redshift are offset slightly along the horizontal axis for legibility; all three series are measured at the same six redshifts. The dashed line marks $b=1$, corresponding to an unbiased tracer, and all panels share a common vertical scale. Uncertainties are propagated from the jackknife; for the massive-halo bias in the $J=100$ configurations they are the larger of the propagated jackknife uncertainty and the draw-to-draw scatter. All three biases rise with redshift, and the uncertainty on the massive-halo bias grows by more than an order of magnitude from the largest massive sample to the smallest.}
    \label{fig:bias}
\end{figure*}

By construction $b_{\mathrm{CCF}}^2 = b_{\mathrm{MMH}}\,b_{\mathrm{FH}}$. The cross-correlation bias is the geometric mean of the other two and cannot fall outside them, so the physical content of the ordering $b_{\mathrm{MMH}} > b_{\mathrm{CCF}} > b_{\mathrm{FH}}$ is that the massive haloes trace rarer peaks than the field.

Figure~\ref{fig:bias} shows all three against redshift. The field bias rises from $0.80$ at $z = 0$ to $4.64$ at $z = 5.29$, the massive-halo bias from $1.74$ to $7.56$ for the largest sample. At $z = 0$ the field bias lies below unity in all three configurations, at $0.80$, $0.81$ and $0.76$. A field dominated by haloes near $10^{11}\,h^{-1}\,\mathrm{M_\odot}$ is less strongly clustered than the mass. The three values agree within their uncertainties of $\pm0.01$, $\pm0.03$ and $\pm0.06$, showing no systematic dependence on the size of the field sample.

The precision with which the massive-halo bias is determined varies sharply between configurations. It is measured to between $\pm 0.04$ and $\pm 0.21$ across the redshift range for the 10000-halo sample, $\pm 0.22$ and $\pm 0.84$ for 500 haloes, and $\pm 1.22$ and $\pm 4.23$ for 50. Precision degrades by more than an order of magnitude between the largest and smallest samples, and Section~\ref{sec:precision_sample_size} quantifies that dependence across all sample and field sizes.

%----------------------------------
%----------------------------------
\subsection{Halo mass recovery}
\label{sec:mass_recovery}

\begin{figure*}
    \centering
    \includegraphics[width=\textwidth]{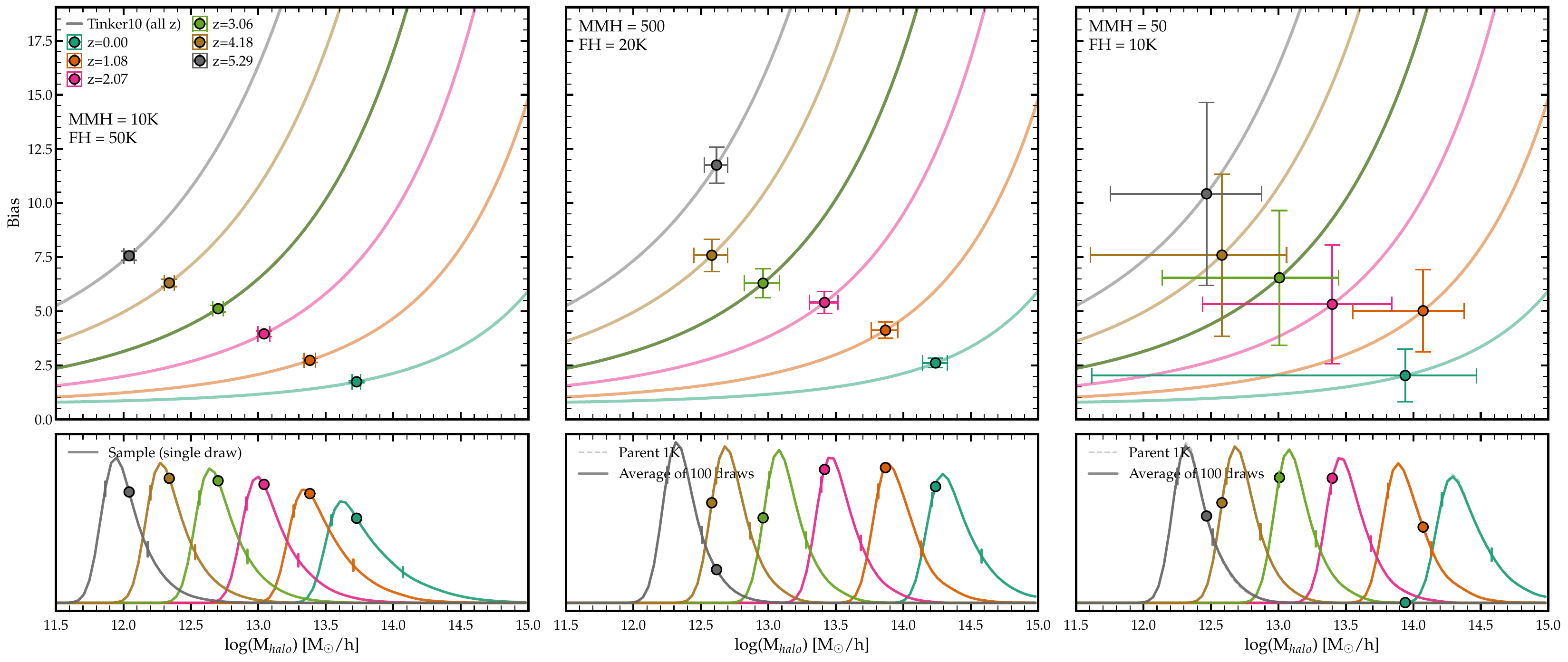}
    \caption{Halo mass recovery for the three standard configurations---best, typical and poor: 10000 massive haloes against a 50000-halo field, 500 against 20000, and 50 against 10000. Top row: the Tinker et al.\ (2010) bias--mass relation at each of the six redshifts, coloured by redshift, with the measured bias and inferred mass plotted as points carrying the bias uncertainty and the asymmetric mass uncertainty. The points lie on their own redshift's curve by construction, since the mass is obtained by inverting that relation. Bottom row: the halo mass distributions corresponding to each point. The left panel shows the single deterministic sample, with $1\sigma$ ticks showing the width of that sample, while the $J=100$ panels show the average of the 100 draws (solid) of the 1000-halo parent pool from which they were drawn, with $1\sigma$ ticks showing the width of the parent-pool distribution and the recovered mass marked by a filled circle. All panels in the top row share a common vertical scale. The bias--mass relation is nearly flat across the mass range occupied by the $z=0$ samples and steepens towards higher redshift, so the inversion is better conditioned there.}
    \label{fig:recovery}
\end{figure*}

A measured bias corresponds to a single mass on the \citet{Tinker2010} relation at each redshift. Figure~\ref{fig:recovery} shows where the three configurations land. The points sit on their own redshift's curve by construction, so the upper panels display the measurement rather than test it; the test is the comparison against the true mass of the population, shown in the lower panels and tabulated in Table~\ref{tab:4mass_j1} and Table~\ref{tab:5mass_j100}.

Two different quantities are being compared. The inversion returns a characteristic mass, the mass whose bias on that relation equals the bias measured for the sample. The true mass is the arithmetic mean of $\log_{10} M_{200c}$ measured directly from the simulation, taken over the haloes of the sample itself for the deterministic selections and over the 1000-halo parent pool for the subsampled ones. These need not agree, because the bias--mass relation is non-linear across the width of the distribution, and the offset $\Delta_{\mathrm{mean}} = \log M_{\mathrm{inf}} - \log (M_{\mathrm{halo}})_{\mathrm{mean}}$ carries that difference alongside the measurement error. An analogous offset against the median, $\Delta_{\mathrm{median}}$, is also shown.

The 10000-halo sample against the 50000-halo field recovers to $|\Delta_{\rm mean}| = 0.079$ dex at $z = 0$, falling to $0.019$ dex by $z = 5.29$. The 500-halo sample against 20000 stays between $0.105$ and $0.236$ dex, while the 50-halo sample against 10000 spans $0.087$ to $0.457$ dex.

For the well-sampled configurations the offset is smaller than the spread of the population it describes. At $z = 0$ the 10000 most massive haloes span $0.266$ dex in log mass about their mean, and the clustering measurement reproduces that mean to $0.079$ dex. This does not hold everywhere. The 50-halo configuration at $z = 0$ is offset by $0.457$ dex from a parent pool only $0.185$ dex wide.

For the $J = 100$ samples the averaged draw distributions reproduce the parent 1000-halo pool, while the $J = 1$ samples are their own top-$M$ selections and therefore have no parent comparison curve.
%--------------------------------------------
%--------------------------------------------

\subsection{Accuracy of the recovered masses}
\label{sec:recovery_accuracy}

\begin{figure*}
    \centering
    \includegraphics[width=\textwidth]{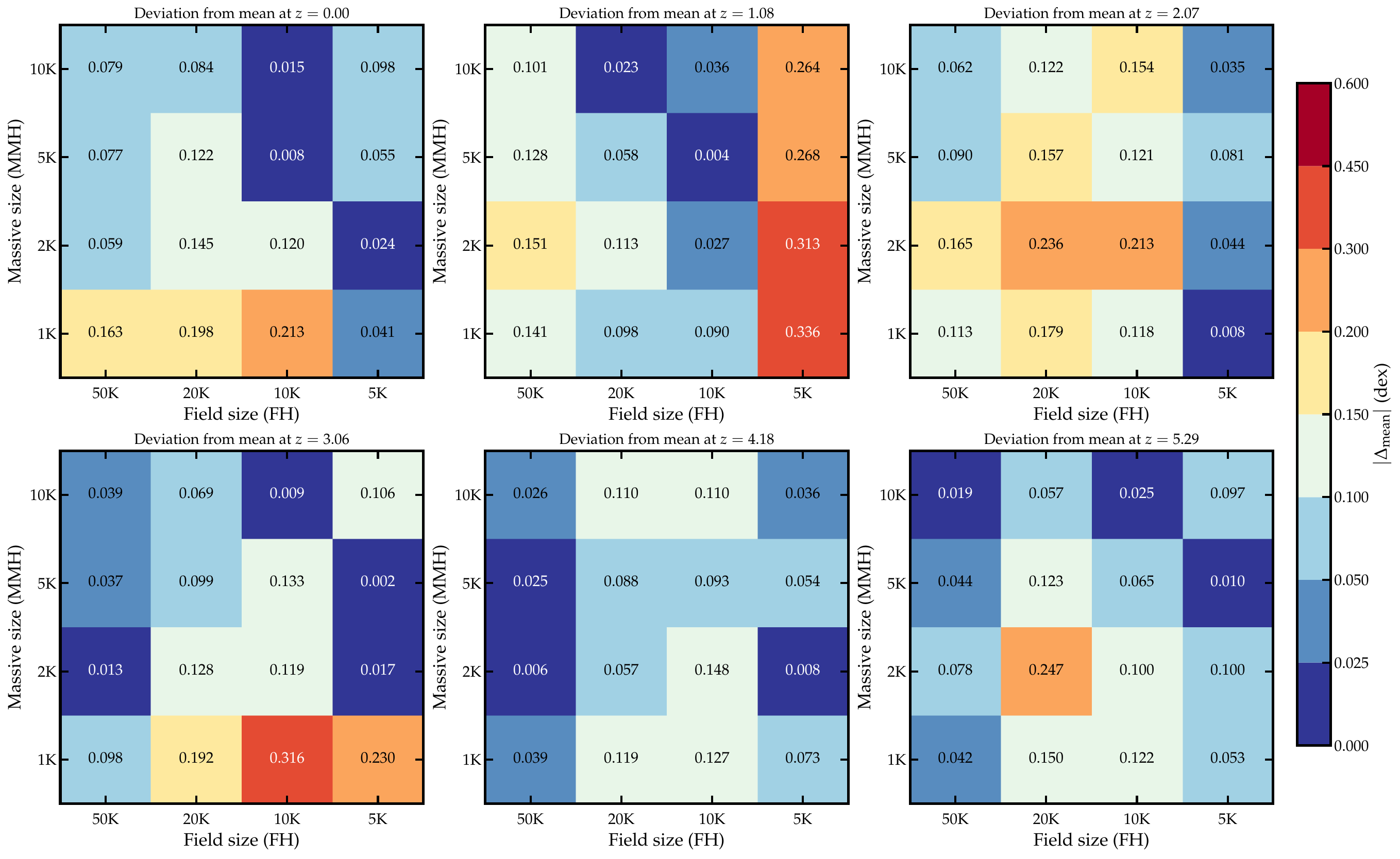}
    \caption{Accuracy of the recovered mass for the deterministic ($J=1$) samples, $|\Delta_{\rm mean}| = |\log(M_{\rm inf}) - \log(M_{\rm halo})_{\rm mean}|$ in dex, as a function of massive sample size (rows) and field sample size (columns) at each of the six redshifts. Values are printed in each cell. The colour scale is fixed between 0 and 0.30 dex and is shared between all panels and with Figure~\ref{fig:J100}, so panels may be compared both with one another and between figures. Offsets tend to be smaller at higher redshift, although the dependence on massive sample size is weak and not ordered. The 5000-halo field at $z=1.08$ stands out as a column of large offsets at every massive sample size.}
    \label{fig:J1}
\end{figure*}

\begin{figure*}
    \centering
    \includegraphics[width=\textwidth]{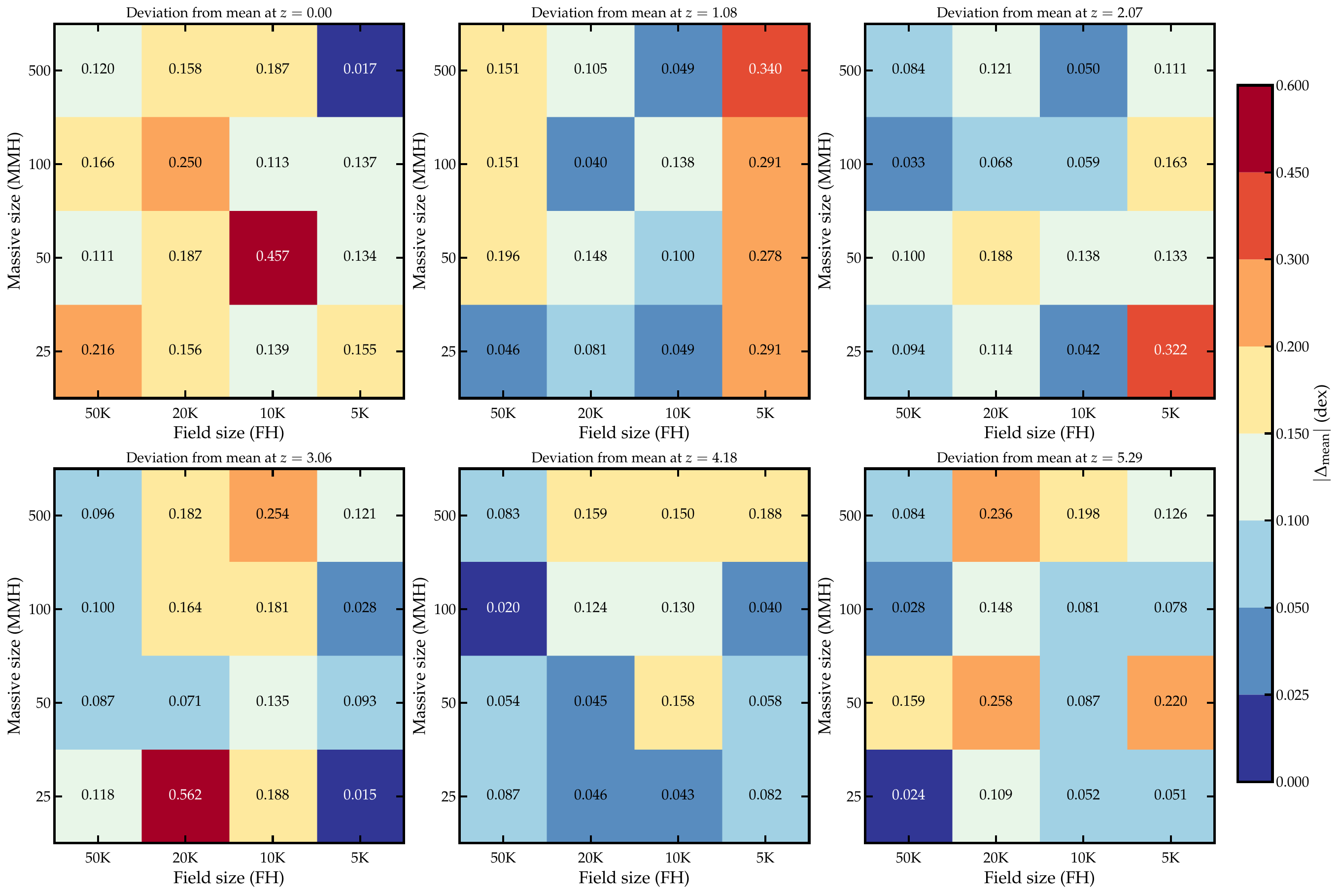}
    \caption{As Figure~\ref{fig:J1}, for the subsampled ($J=100$) samples, where each measurement is the average of 100 random draws from the 1000 most massive haloes; here $\rm log(M_{\rm halo})_{\rm mean}$ is that of 1000-halo parent pool. The colour scale is identical to that of Figure~\ref{fig:J1}. The column of large offsets at $z=1.08$ for the 5000-halo field persists across all four massive sample sizes.}
    \label{fig:J100}
\end{figure*}

The two heat maps in Figures~\ref{fig:J1} and~\ref{fig:J100} give $|\Delta_{\rm mean}|$ for all 192 configurations, on a colour scale shared between panels and between figures. The median offset is $0.103$ dex. Sixty-eight per cent of configurations fall within $0.136$ dex, ninety per cent within $0.219$ dex, and the largest is $0.562$ dex, their values are printed in the cells.

Median offsets run from $0.065$ dex at 10000 haloes to $0.134$ dex at 50, but not in sequence: the 25-halo samples return $0.091$ dex, comparable to samples twenty times larger. Recovery accuracy is not determined by sample size alone.

Redshift orders more cleanly. The median offset falls from $0.128$ dex at $z = 0$ to $0.078$ dex at $z = 4.18$, rising slightly to $0.086$ dex at $z = 5.29$. Figure~\ref{fig:recovery} shows why. At $z = 0$ the bias--mass relation is nearly flat across the range the massive samples occupy, turning upward only at the highest masses, so the inversion is poorly conditioned there. At higher redshift the relation is steep and uniformly curved over the same range. The samples narrow as well, their spread about the mean falling from $0.266$ to $0.158$ dex for the 10000-halo samples (Table~\ref{tab:4mass_j1}). Both effects reduce the offset introduced by the non-linearity discussed in Section~\ref{sec:mass_recovery}.

The sign of the offset varies systematically with redshift. At $z = 0$, 84 per cent of configurations under-recover the true mass, and the mean signed offset is $-0.118$ dex. Under-recovery persists through $z = 4.18$, and only at $z = 5.29$ does the majority over-recover, with a mean of $+0.069$ dex. The apparently positive mean at $z = 1.08$ is driven by the anomalous 5000-halo field discussed in Section~\ref{sec:field_reference}; excluding that field, the mean there is $-0.054$ dex with 75 per cent of configurations under-recovering. The signed offsets for every configuration are listed in the Supporting Information.

%--------------------------------------------
%--------------------------------------------

\subsection{Precision and sample size}
\label{sec:precision_sample_size}

The uncertainty on the recovered mass grows monotonically as the sample shrinks: $\pm 0.081$ dex at 10000 haloes, $\pm 0.089$ at 5000, $\pm 0.118$ at 2000, $\pm 0.132$ at 1000, $\pm 0.134$ at 500, $\pm 0.311$ at 100, $\pm 0.477$ at 50 and $\pm 0.709$ at 25. These are medians over all four field sizes and six redshifts; Table~\ref{tab:4mass_j1} and Table~\ref{tab:5mass_j100} list a subset of configurations, and the complete set of asymmetric bounds is given in the supplementary material. They are measurement uncertainties and exclude the calibration scatter of the bias--mass relation itself, which is discussed in Section~\ref{sec:sample_size}.

The bounds are asymmetric throughout. At 25 haloes the median lower bound extends $0.973$ dex and the upper $0.457$ dex. The bias--mass relation flattens towards lower mass, so a downward excursion in bias maps to a larger excursion in mass than an upward one of the same size. At 10000 haloes the median lower and upper bounds are $0.084$ and $0.077$ dex, respectively, so the asymmetry has almost vanished.

The return on enlarging a sample is not uniform. Doubling from 25 to 50 haloes reduces the uncertainty by 33 per cent, and from 50 to 100 by a further 35 per cent. Within the deterministic samples the gains are much smaller: 11 per cent from 1000 to 2000 and a further 10 per cent from 5000 to 10000. The 500 and 1000-halo samples are not directly comparable, since the former is drawn from the 1000-halo parent pool (J = 100) while the latter is deterministic (J = 1).

However, the comparison across 500 to 1000 haloes crosses a change of estimator. Samples of 1000 and above are single draws, while those of 500 and below are averages of 100 draws, so the 1.5 per cent step is not a purely physical result. Within each regime the pattern holds: large returns at the smallest sample sizes, modest returns above 1000.

Sample size affects accuracy and precision differently. Precision improves with size steeply and in sequence. Accuracy improves only weakly and not in sequence, because a second effect works against it. A larger sample reaches further down the mass function, so the population that a single characteristic mass must represent grows broader: at $z = 0$ the 10000 most massive haloes span $0.266$ dex about their mean, against $0.239$ dex for 5000 and $0.185$ dex for the 1000-halo pool from which the smaller samples are drawn (Tables~\ref{tab:4mass_j1} and \ref{tab:5mass_j100}). The non-linearity discussed in Section~\ref{sec:mass_recovery} acts across that wider range. Larger samples are measured better and describe a less homogeneous population, and the two effects partly offset, which is why the ordering in Section~\ref{sec:recovery_accuracy} is imperfect.

%-------------------------------------------------
%-------------------------------------------------

\subsection{Error from the reference sample}
\label{sec:field_reference}

The field sample is a reference rather than the population of interest, but an error in it does not stay there. Every cross-correlation at a given redshift is measured against the same field, so an error in the field auto-correlation displaces every mass inferred from it in the same direction and by a similar amount. This section isolates that displacement, which is accessible here only because the true masses are known.

One realisation makes the effect visible. At $z = 1.08$ the 5000-halo field returns $\xi(8) = 0.161 \pm 0.089$, against $0.266$ to $0.296$ for the three larger fields at the same epoch (Table~\ref{tab:1acf}), a deficit of some 45 per cent and comparable to the uncertainty on the measurement itself. Because the massive-halo bias varies inversely with the square root of the field auto-correlation (Equations~\ref{eq:bias_acf} and~\ref{eq:bias_massive}), the inferred bias rises accordingly: for the 1000-halo massive sample at $z = 1.08$, from $3.99$ against the 50000-halo field to $6.42$ against the 5000-halo field (Table~\ref{tab:4mass_j1}). Every massive sample paired with that field is displaced together, by $+0.26$ to $+0.34$ dex across all eight massive-sample sizes, from 25 haloes to 10000 (Table~\ref{tab:4mass_j1} and \ref{tab:5mass_j100}). The field mass distribution is otherwise consistent with those of the larger reference samples, so the displacement originates in the measured clustering amplitude rather than an anomalous halo selection.

The displacement is therefore a property of the reference, not of the rare population, and no increase in the number of rare objects reduces it. Four field sizes at six redshifts cannot locate the threshold below which this becomes the limiting term, but they establish that reference-sample variance contributes to the error budget independently of the rare-object sample.

%=================================
% TABLE 1 -- field auto-correlations (ACF).
%=================================
\begin{table}
\centering
\small
\setlength{\tabcolsep}{8pt}
\renewcommand{\arraystretch}{1.08}
\caption{Correlation function parameters for the field--field auto-correlations (ACF), measured in the Millennium simulation over $0 \leq z \leq 5.29$ for field samples of 50000, 20000, 10000 and 5000 haloes. $\xi_8$ is the measured correlation amplitude at $r = 8\,h^{-1}\,\mathrm{Mpc}$, with 27-subvolume jackknife uncertainty; $r_0$ and $\gamma$ are the correlation length and slope of the fitted power-law model, $\xi(r) = (r/r_0)^{-\gamma}$. One field sample is drawn per field size and redshift and reused for every massive sample, so each field measurement is shared by all corresponding configurations in Tables~\ref{tab:2ccf_j1} and~\ref{tab:3ccf_j100}.}
\label{tab:1acf}
\begin{tabular}{rccc}
\cmidrule(l{6pt}r{6pt}){1-4}
Field & $\xi_8$ & $r_0\,[h^{-1}\,{\rm Mpc}]$ & $\gamma$\\
\cmidrule(l{6pt}r{6pt}){1-4}
\multicolumn{4}{c}{ACF: $z = 0.0000$ (Age $= 13.58$\,Gyr)}\\
\cmidrule(l{6pt}r{6pt}){1-4}
50000 & $0.399 \pm 0.015$ & 4.38 & 1.52\\
20000 & $0.410 \pm 0.029$ & 4.38 & 1.56\\
10000 & $0.361 \pm 0.053$ & 4.61 & 1.62\\
5000 & $0.309 \pm 0.089$ & 4.65 & 1.64\\
\cmidrule(l{6pt}r{6pt}){1-4}
\multicolumn{4}{c}{ACF: $z = 1.0779$ (Age $= 5.67$\,Gyr)}\\
\cmidrule(l{6pt}r{6pt}){1-4}
50000 & $0.296 \pm 0.011$ & 3.31 & 1.48\\
20000 & $0.293 \pm 0.024$ & 3.35 & 1.53\\
10000 & $0.266 \pm 0.059$ & 3.49 & 1.62\\
5000 & $0.161 \pm 0.088$ & 4.13 & 1.60\\
\cmidrule(l{6pt}r{6pt}){1-4}
\multicolumn{4}{c}{ACF: $z = 2.0700$ (Age $= 3.27$\,Gyr)}\\
\cmidrule(l{6pt}r{6pt}){1-4}
50000 & $0.300 \pm 0.010$ & 3.28 & 1.49\\
20000 & $0.328 \pm 0.024$ & 3.45 & 1.57\\
10000 & $0.342 \pm 0.041$ & 3.45 & 1.62\\
5000 & $0.304 \pm 0.105$ & 3.46 & 1.52\\
\cmidrule(l{6pt}r{6pt}){1-4}
\multicolumn{4}{c}{ACF: $z = 3.0604$ (Age $= 2.17$\,Gyr)}\\
\cmidrule(l{6pt}r{6pt}){1-4}
50000 & $0.335 \pm 0.010$ & 3.78 & 1.52\\
20000 & $0.365 \pm 0.031$ & 3.81 & 1.52\\
10000 & $0.369 \pm 0.057$ & 4.03 & 1.56\\
5000 & $0.326 \pm 0.111$ & 4.59 & 1.53\\
\cmidrule(l{6pt}r{6pt}){1-4}
\multicolumn{4}{c}{ACF: $z = 4.1795$ (Age $= 1.51$\,Gyr)}\\
\cmidrule(l{6pt}r{6pt}){1-4}
50000 & $0.459 \pm 0.012$ & 4.62 & 1.57\\
20000 & $0.469 \pm 0.024$ & 4.63 & 1.59\\
10000 & $0.467 \pm 0.053$ & 4.75 & 1.66\\
5000 & $0.432 \pm 0.083$ & 4.82 & 1.63\\
\cmidrule(l{6pt}r{6pt}){1-4}
\multicolumn{4}{c}{ACF: $z = 5.2888$ (Age $= 1.13$\,Gyr)}\\
\cmidrule(l{6pt}r{6pt}){1-4}
50000 & $0.612 \pm 0.021$ & 5.75 & 1.67\\
20000 & $0.580 \pm 0.031$ & 5.62 & 1.69\\
10000 & $0.645 \pm 0.061$ & 5.96 & 1.76\\
5000 & $0.762 \pm 0.104$ & 6.28 & 1.70\\
\cmidrule(l{6pt}r{6pt}){1-4}
\end{tabular}
\end{table}

%=================================
% TABLE 2 -- CCF, deterministic samples (J = 1).
%=================================
\begin{table}
\centering
\small
\setlength{\tabcolsep}{7pt}
\renewcommand{\arraystretch}{1.08}
\caption{Correlation function parameters for the massive--field cross-correlations (CCF) of the deterministic samples ($J=1$), measured over $0 \leq z \leq 5.29$. The main-text table shows the representative configuration of 10000 massive haloes against a 50000-halo field at all six redshifts; the complete set of configurations is provided in the supplementary material. Symbols are as in Table~\ref{tab:1acf}, and the quoted uncertainty on $\xi_8$ is the 27-subvolume jackknife uncertainty.}
\label{tab:2ccf_j1}
\begin{tabular}{rrrrcc}
\cmidrule(l{6pt}r{6pt}){1-6}
$z$ & Massive & Field & $\xi_8$ & $r_0$ & $\gamma$\\
\cmidrule(l{6pt}r{6pt}){1-6}
0.0000 & 10000 & 50000 & $0.870 \pm 0.025$ & 7.22 & 1.83\\
1.0779 & 10000 & 50000 & $0.721 \pm 0.020$ & 6.25 & 1.66\\
2.0700 & 10000 & 50000 & $0.736 \pm 0.028$ & 6.19 & 1.57\\
3.0604 & 10000 & 50000 & $0.769 \pm 0.026$ & 6.56 & 1.58\\
4.1795 & 10000 & 50000 & $0.873 \pm 0.026$ & 7.32 & 1.67\\
5.2888 & 10000 & 50000 & $0.998 \pm 0.037$ & 7.97 & 1.79\\
\cmidrule(l{6pt}r{6pt}){1-6}
\end{tabular}
\end{table}

%=================================
% TABLE 3 -- CCF, subsampled samples (J = 100).
%=================================
\begin{table}
\centering
\small
\setlength{\tabcolsep}{7pt}
\renewcommand{\arraystretch}{1.08}
\caption{As Table~\ref{tab:2ccf_j1}, for the subsampled massive samples ($J=100$). The main-text table shows the representative configurations of 500 massive haloes against a 20000-halo field and 50 massive haloes against a 10000-halo field at all six redshifts; the complete set of configurations is provided in the supplementary material. For each configuration, the correlation functions of 100 independent random draws from the 1000 most massive haloes are averaged before the correlation parameters are measured.}
\label{tab:3ccf_j100}

\begin{tabular}{rrrrcc}
\cmidrule(l{6pt}r{6pt}){1-6}
$z$ & Massive & Field & $\xi_8$ & $r_0$ & $\gamma$\\
\cmidrule(l{6pt}r{6pt}){1-6}

0.0000 & 500 & 20000 & $1.321 \pm 0.117$ & 9.31 & 1.96\\
1.0779 & 500 & 20000 & $1.082 \pm 0.094$ & 8.09 & 1.73\\
2.0700 & 500 & 20000 & $1.052 \pm 0.090$ & 7.73 & 1.68\\
3.0604 & 500 & 20000 & $0.987 \pm 0.096$ & 8.02 & 1.60\\
4.1795 & 500 & 20000 & $1.061 \pm 0.113$ & 8.51 & 1.74\\
5.2888 & 500 & 20000 & $1.508 \pm 0.108$ & 9.47 & 1.88\\

\cmidrule(l{6pt}r{6pt}){1-6}

0.0000 & 50 & 10000 & $0.965 \pm 0.580$ & 9.43 & 1.99\\
1.0779 & 50 & 10000 & $1.256 \pm 0.475$ & 8.80 & 1.64\\
2.0700 & 50 & 10000 & $1.057 \pm 0.545$ & 7.92 & 1.76\\
3.0604 & 50 & 10000 & $1.032 \pm 0.490$ & 8.30 & 1.58\\
4.1795 & 50 & 10000 & $1.060 \pm 0.522$ & 8.93 & 1.69\\
5.2888 & 50 & 10000 & $1.412 \pm 0.573$ & 9.86 & 1.93\\

\cmidrule(l{6pt}r{6pt}){1-6}
\end{tabular}
\end{table}
\section{Discussion}
\label{sec:discussion}

\subsection{Competing effects of sample size}
\label{sec:sample_size}

Across all 192 configurations the median offset between the inferred and true mass is 0.103 dex, with 68 per cent of configurations within 0.136 dex. That figure depends only weakly on how many massive objects were used. The uncertainty depends on it steeply, from $\pm0.081$ dex at 10000 haloes to $\pm0.709$ dex at 25.

The insensitivity of the offset follows from the inversion. A measured bias is the effective bias of the population, a number-weighted average of halo bias over the sample \citep{Mo1996,Desjacques2018}, and inverting it gives the mass whose bias equals that average. That this is not the sample's mean halo mass is expected, since the relation is non-linear across the width of any real sample. What has not been measurable in observational work is the size of the difference, the true masses being inaccessible. Here it can be measured directly, though the value obtained is specific to this simulation and to the bias relation adopted. A larger sample reaches further down the mass function and the population that a single characteristic mass must represent grows broader, which increases the potential discrepancy.

The departure from linearity can be isolated. Computing a bias and a mass for each of the 100 draws separately and averaging afterwards returns exactly the same massive-halo bias as averaging the correlation functions first, since that bias is linear in the cross-correlation amplitude and the field auto-correlation is common to every draw. The inferred masses from the two routes differ, by less than the quoted uncertainties, and that difference is a direct measure of the non-linearity of the inversion across the spread of the individual draws.

An additional systematic is set by the calibration of the bias--mass relation and is not reduced by increasing the sample size. \citet{Tinker2010} find an approximately 6 per cent scatter about their best-fitting halo bias relation. The logarithmic slope $\mathrm{d}\log b/\mathrm{d}\log M$ runs between 0.31 and 0.43 across the samples and redshifts considered here, so that scatter corresponds to between 0.060 and 0.082 dex in inferred mass, with a median of 0.069 dex. Our best configurations reach $\pm0.081$ dex from the clustering measurement alone, comparable to that systematic. Treating the two as independent and combining them in quadrature gives approximately 0.11 dex. Further gains in statistical precision would increasingly expose the calibration uncertainty rather than remove it.

%-------------------------------------------------
%-------------------------------------------------

\subsection{How many objects are enough}
\label{sec:sample_size_requirements}

An observer holding a handful of rare objects faces a practical question: whether more telescope time would materially improve the answer. Section~\ref{sec:precision_sample_size} answers it directly. Doubling a sample from

%-------------------------------------------------
%-------------------------------------------------

%-------------------------------------------------
%-------------------------------------------------
% to feel the blanck page before and after the table CF
\clearpage
\onecolumn

% Horizontal spacing between columns
\setlength{\tabcolsep}{12pt}

% Vertical spacing between rows
\renewcommand{\arraystretch}{1.15}

% Caption spans the full table/text width
\setlength{\LTcapwidth}{\textwidth}

% ============================================================
% TABLE 4: J = 1
% ============================================================

\begin{longtable}{ccccccc}
\caption{Halo mass recovery for the deterministic massive samples ($J=1$), over $0 \leq z \leq 5.29$. The main-text table lists the representative 10000-halo sample against the 50000-halo field, the 1000-halo sample against the 50000-halo field, and all four massive-sample sizes (10000, 5000, 2000 and 1000) against the 5000-halo field; the complete set of configurations is provided in the supplementary material. For each configuration the massive-halo bias is listed with its uncertainty, alongside the inferred mass, $\log(M_{\rm inf})$, obtained by inverting the bias--mass relation, with asymmetric bounds. $\log(M_{\rm halo})_{\rm mean}$ and $\log(M_{\rm halo})_{\rm median}$ are the mean and median of $\log_{10} M_{200c}$ measured directly from the halo mass distribution of the sample in the simulation; the $1\sigma$ width quoted with the mean and the percentile bounds quoted with the median describe that distribution rather than the uncertainty on either quantity. The offsets $\Delta_{\rm mean} = \log(M_{\rm inf}) - \log(M_{\rm halo})_{\rm mean}$ and $\Delta_{\rm median} = \log(M_{\rm inf}) - \log(M_{\rm halo})_{\rm median}$ quantify how well the bias-inferred mass reproduces each, with positive values indicating an overestimate and negative values an underestimate. Quoted uncertainties on the bias and on $\log(M_{\rm inf})$ are measurement uncertainties and exclude the calibration scatter of the bias--mass relation itself, discussed in Section~\ref{sec:sample_size}. All masses are in $h^{-1}\,M_\odot$.}
\label{tab:4mass_j1}\\
\hline
$z$ & Bias & $\log(M_{\rm inf})$ & $\log(M_{\rm halo})_{\rm mean}$ &
$\Delta_{\rm mean}$ & $\log(M_{\rm halo})_{\rm median}$ &
$\Delta_{\rm median}$\\
\hline
\endfirsthead

\multicolumn{7}{l}{\textit{Table \ref{tab:mass_j1} -- continued}}\\
\hline
$z$ & Bias & $\log(M_{\rm inf})$ & $\log(M_{\rm halo})_{\rm mean}$ &
$\Delta_{\rm mean}$ & $\log(M_{\rm halo})_{\rm median}$ &
$\Delta_{\rm median}$\\
\hline
\endhead

\hline
\endfoot

\hline
\multicolumn{7}{c}{MMH $= 10000$, Field $= 50000$}\\
\hline
0.0000 & $1.740^{+0.037}_{-0.037}$ & $13.728^{+0.031}_{-0.032}$ & $13.807 \pm 0.266$ & $-0.079$ & $13.733^{+0.329}_{-0.168}$ & $-0.005$\\
1.0779 & $2.733^{+0.087}_{-0.087}$ & $13.382^{+0.040}_{-0.042}$ & $13.483 \pm 0.220$ & $-0.101$ & $13.423^{+0.269}_{-0.141}$ & $-0.041$\\
2.0700 & $3.955^{+0.141}_{-0.141}$ & $13.042^{+0.044}_{-0.046}$ & $13.104 \pm 0.194$ & $-0.062$ & $13.049^{+0.238}_{-0.118}$ & $-0.007$\\
3.0604 & $5.118^{+0.151}_{-0.151}$ & $12.702^{+0.038}_{-0.039}$ & $12.741 \pm 0.178$ & $-0.039$ & $12.691^{+0.215}_{-0.107}$ & $0.011$\\
4.1795 & $6.310^{+0.170}_{-0.170}$ & $12.340^{+0.036}_{-0.037}$ & $12.366 \pm 0.165$ & $-0.026$ & $12.318^{+0.201}_{-0.099}$ & $0.022$\\
5.2888 & $7.563^{+0.211}_{-0.211}$ & $12.042^{+0.038}_{-0.040}$ & $12.023 \pm 0.158$ & $0.019$ & $11.977^{+0.192}_{-0.092}$ & $0.065$\\
\hline

\multicolumn{7}{c}{MMH $= 1000$, Field $= 50000$}\\
\hline
0.0000 & $2.600^{+0.168}_{-0.168}$ & $14.235^{+0.068}_{-0.075}$ & $14.397 \pm 0.185$ & $-0.163$ & $14.347^{+0.226}_{-0.123}$ & $-0.112$\\
1.0779 & $3.989^{+0.268}_{-0.268}$ & $13.832^{+0.070}_{-0.077}$ & $13.973 \pm 0.159$ & $-0.141$ & $13.929^{+0.178}_{-0.100}$ & $-0.098$\\
2.0700 & $5.443^{+0.348}_{-0.348}$ & $13.425^{+0.069}_{-0.075}$ & $13.538 \pm 0.147$ & $-0.113$ & $13.492^{+0.184}_{-0.087}$ & $-0.067$\\
3.0604 & $6.750^{+0.485}_{-0.485}$ & $13.044^{+0.081}_{-0.089}$ & $13.143 \pm 0.139$ & $-0.098$ & $13.102^{+0.174}_{-0.085}$ & $-0.057$\\
4.1795 & $8.349^{+0.379}_{-0.379}$ & $12.700^{+0.054}_{-0.058}$ & $12.739 \pm 0.126$ & $-0.039$ & $12.708^{+0.156}_{-0.087}$ & $-0.008$\\
5.2888 & $10.055^{+0.562}_{-0.562}$ & $12.421^{+0.069}_{-0.074}$ & $12.380 \pm 0.128$ & $0.042$ & $12.340^{+0.168}_{-0.071}$ & $0.082$\\
\hline

\multicolumn{7}{c}{MMH $= 10000$, Field $= 5000$}\\
\hline
0.0000 & $1.977^{+0.323}_{-0.323}$ & $13.905^{+0.189}_{-0.253}$ & $13.807 \pm 0.266$ & $0.098$ & $13.733^{+0.329}_{-0.168}$ & $0.172$\\
1.0779 & $3.695^{+1.084}_{-1.084}$ & $13.747^{+0.276}_{-0.425}$ & $13.483 \pm 0.220$ & $0.264$ & $13.423^{+0.269}_{-0.141}$ & $0.323$\\
2.0700 & $4.274^{+0.819}_{-0.819}$ & $13.139^{+0.209}_{-0.274}$ & $13.104 \pm 0.194$ & $0.035$ & $13.049^{+0.238}_{-0.118}$ & $0.090$\\
3.0604 & $5.738^{+0.947}_{-0.947}$ & $12.847^{+0.186}_{-0.232}$ & $12.741 \pm 0.178$ & $0.106$ & $12.691^{+0.215}_{-0.107}$ & $0.156$\\
4.1795 & $6.610^{+0.861}_{-0.861}$ & $12.402^{+0.159}_{-0.190}$ & $12.366 \pm 0.165$ & $0.036$ & $12.318^{+0.201}_{-0.099}$ & $0.084$\\
5.2888 & $6.968^{+0.753}_{-0.753}$ & $11.926^{+0.145}_{-0.167}$ & $12.023 \pm 0.158$ & $-0.097$ & $11.977^{+0.192}_{-0.092}$ & $-0.051$\\
\hline

\multicolumn{7}{c}{MMH $= 5000$, Field $= 5000$}\\
\hline
0.0000 & $2.238^{+0.371}_{-0.371}$ & $14.062^{+0.177}_{-0.234}$ & $14.007 \pm 0.239$ & $0.055$ & $13.941^{+0.294}_{-0.153}$ & $0.121$\\
1.0779 & $4.312^{+1.287}_{-1.287}$ & $13.915^{+0.264}_{-0.404}$ & $13.648 \pm 0.199$ & $0.268$ & $13.592^{+0.246}_{-0.127}$ & $0.323$\\
2.0700 & $5.002^{+0.864}_{-0.864}$ & $13.328^{+0.180}_{-0.229}$ & $13.247 \pm 0.179$ & $0.081$ & $13.196^{+0.218}_{-0.108}$ & $0.132$\\
3.0604 & $5.840^{+1.092}_{-1.092}$ & $12.869^{+0.207}_{-0.266}$ & $12.871 \pm 0.166$ & $-0.002$ & $12.821^{+0.203}_{-0.097}$ & $0.048$\\
4.1795 & $7.356^{+1.004}_{-1.004}$ & $12.541^{+0.161}_{-0.193}$ & $12.487 \pm 0.154$ & $0.054$ & $12.442^{+0.189}_{-0.094}$ & $0.099$\\
5.2888 & $8.051^{+0.945}_{-0.945}$ & $12.128^{+0.149}_{-0.174}$ & $12.139 \pm 0.148$ & $-0.010$ & $12.096^{+0.179}_{-0.088}$ & $0.033$\\
\hline

\multicolumn{7}{c}{MMH $= 2000$, Field $= 5000$}\\
\hline
0.0000 & $2.674^{+0.586}_{-0.586}$ & $14.265^{+0.205}_{-0.290}$ & $14.241 \pm 0.207$ & $0.024$ & $14.188^{+0.253}_{-0.139}$ & $0.078$\\
1.0779 & $5.459^{+1.706}_{-1.706}$ & $14.155^{+0.254}_{-0.391}$ & $13.842 \pm 0.175$ & $0.313$ & $13.793^{+0.220}_{-0.111}$ & $0.362$\\
2.0700 & $5.217^{+1.181}_{-1.181}$ & $13.376^{+0.226}_{-0.309}$ & $13.420 \pm 0.159$ & $-0.044$ & $13.373^{+0.199}_{-0.096}$ & $0.003$\\
3.0604 & $6.578^{+1.606}_{-1.606}$ & $13.014^{+0.251}_{-0.350}$ & $13.031 \pm 0.150$ & $-0.017$ & $12.986^{+0.189}_{-0.089}$ & $0.028$\\
4.1795 & $7.874^{+1.383}_{-1.383}$ & $12.628^{+0.198}_{-0.250}$ & $12.636 \pm 0.138$ & $-0.008$ & $12.594^{+0.172}_{-0.083}$ & $0.033$\\
5.2888 & $9.734^{+1.481}_{-1.481}$ & $12.380^{+0.178}_{-0.218}$ & $12.280 \pm 0.136$ & $0.100$ & $12.238^{+0.164}_{-0.075}$ & $0.142$\\
\hline

\multicolumn{7}{c}{MMH $= 1000$, Field $= 5000$}\\
\hline
0.0000 & $2.913^{+0.762}_{-0.762}$ & $14.357^{+0.230}_{-0.344}$ & $14.397 \pm 0.185$ & $-0.041$ & $14.347^{+0.226}_{-0.123}$ & $0.010$\\
1.0779 & $6.422^{+2.113}_{-2.113}$ & $14.309^{+0.252}_{-0.394}$ & $13.973 \pm 0.159$ & $0.336$ & $13.929^{+0.178}_{-0.100}$ & $0.379$\\
2.0700 & $6.072^{+1.540}_{-1.540}$ & $13.546^{+0.237}_{-0.335}$ & $13.538 \pm 0.147$ & $0.008$ & $13.492^{+0.184}_{-0.087}$ & $0.054$\\
3.0604 & $6.050^{+1.662}_{-1.662}$ & $12.913^{+0.286}_{-0.417}$ & $13.143 \pm 0.139$ & $-0.230$ & $13.102^{+0.174}_{-0.085}$ & $-0.189$\\
4.1795 & $8.119^{+1.735}_{-1.735}$ & $12.666^{+0.234}_{-0.310}$ & $12.739 \pm 0.126$ & $-0.073$ & $12.708^{+0.156}_{-0.087}$ & $-0.043$\\
5.2888 & $10.146^{+2.215}_{-2.215}$ & $12.433^{+0.244}_{-0.325}$ & $12.380 \pm 0.128$ & $0.053$ & $12.340^{+0.168}_{-0.071}$ & $0.093$\\
\hline
\end{longtable}

% ============================================================
% TABLE 5: J = 100
% ============================================================

\begin{longtable}{ccccccc}
\caption{As Table~\ref{tab:4mass_j1}, for the subsampled massive samples ($J=100$). The main-text table lists the representative 500-halo sample against the 20000-halo field, the representative 50-halo sample against the 10000-halo field, and all four massive-sample sizes (500, 100, 50 and 25) against the 5000-halo field; the complete set of configurations is provided in the supplementary material. The bias is measured from the average of 100 independent random draws from the 1000 most massive haloes. Because the samples are random subsamples of a common parent pool, the quoted mean, median, $1\sigma$ width and percentile bounds are all those of the 1000-halo parent distribution at each redshift.}
\label{tab:5mass_j100}\\
\hline
$z$ & Bias & $\log(M_{\rm inf})$ & $\log(M_{\rm halo})_{\rm mean}$ &
$\Delta_{\rm mean}$ & $\log(M_{\rm halo})_{\rm median}$ &
$\Delta_{\rm median}$\\
\hline
\endfirsthead

\multicolumn{7}{l}{\textit{Table \ref{tab:mass_j100} -- continued}}\\
\hline
$z$ & Bias & $\log(M_{\rm inf})$ & $\log(M_{\rm halo})_{\rm mean}$ &
$\Delta_{\rm mean}$ & $\log(M_{\rm halo})_{\rm median}$ &
$\Delta_{\rm median}$\\
\hline
\endhead

\hline
\endfoot

\hline
\multicolumn{7}{c}{MMH $= 500$, Field $= 20000$}\\
\hline
0.0000 & $2.610^{+0.215}_{-0.215}$ & $14.239^{+0.086}_{-0.097}$ & $14.397 \pm 0.185$ & $-0.158$ & $14.347^{+0.226}_{-0.123}$ & $-0.108$\\
1.0779 & $4.124^{+0.375}_{-0.375}$ & $13.868^{+0.092}_{-0.105}$ & $13.973 \pm 0.159$ & $-0.105$ & $13.929^{+0.178}_{-0.100}$ & $-0.062$\\
2.0700 & $5.404^{+0.498}_{-0.498}$ & $13.417^{+0.098}_{-0.111}$ & $13.538 \pm 0.147$ & $-0.121$ & $13.492^{+0.184}_{-0.087}$ & $-0.075$\\
3.0604 & $6.295^{+0.671}_{-0.671}$ & $12.961^{+0.120}_{-0.139}$ & $13.143 \pm 0.139$ & $-0.182$ & $13.102^{+0.174}_{-0.085}$ & $-0.141$\\
4.1795 & $7.585^{+0.748}_{-0.748}$ & $12.580^{+0.118}_{-0.134}$ & $12.739 \pm 0.126$ & $-0.159$ & $12.708^{+0.156}_{-0.087}$ & $-0.128$\\
5.2888 & $11.755^{+0.839}_{-0.839}$ & $12.616^{+0.083}_{-0.091}$ & $12.380 \pm 0.128$ & $0.236$ & $12.340^{+0.168}_{-0.071}$ & $0.276$\\
\hline

\multicolumn{7}{c}{MMH $= 50$, Field $= 10000$}\\
\hline
0.0000 & $2.030^{+1.220}_{-1.220}$ & $13.940^{+0.528}_{-2.324}$ & $14.397 \pm 0.185$ & $-0.457$ & $14.347^{+0.226}_{-0.123}$ & $-0.407$\\
1.0779 & $5.023^{+1.897}_{-1.897}$ & $14.073^{+0.305}_{-0.522}$ & $13.973 \pm 0.159$ & $0.100$ & $13.929^{+0.178}_{-0.100}$ & $0.143$\\
2.0700 & $5.323^{+2.746}_{-2.746}$ & $13.399^{+0.442}_{-0.961}$ & $13.538 \pm 0.147$ & $-0.138$ & $13.492^{+0.184}_{-0.087}$ & $-0.092$\\
3.0604 & $6.546^{+3.108}_{-3.108}$ & $13.008^{+0.437}_{-0.868}$ & $13.143 \pm 0.139$ & $-0.135$ & $13.102^{+0.174}_{-0.085}$ & $-0.094$\\
4.1795 & $7.592^{+3.744}_{-3.744}$ & $12.582^{+0.479}_{-0.976}$ & $12.739 \pm 0.126$ & $-0.158$ & $12.708^{+0.156}_{-0.087}$ & $-0.127$\\
5.2888 & $10.426^{+4.229}_{-4.229}$ & $12.467^{+0.409}_{-0.712}$ & $12.380 \pm 0.128$ & $0.087$ & $12.340^{+0.168}_{-0.071}$ & $0.127$\\
\hline

\multicolumn{7}{c}{MMH $= 500$, Field $= 5000$}\\
\hline
0.0000 & $3.081^{+0.785}_{-0.785}$ & $14.415^{+0.219}_{-0.322}$ & $14.397 \pm 0.185$ & $0.017$ & $14.347^{+0.226}_{-0.123}$ & $0.068$\\
1.0779 & $6.447^{+2.209}_{-2.209}$ & $14.312^{+0.261}_{-0.415}$ & $13.973 \pm 0.159$ & $0.340$ & $13.929^{+0.178}_{-0.100}$ & $0.383$\\
2.0700 & $6.684^{+1.608}_{-1.608}$ & $13.649^{+0.220}_{-0.304}$ & $13.538 \pm 0.147$ & $0.111$ & $13.492^{+0.184}_{-0.087}$ & $0.157$\\
3.0604 & $6.620^{+1.719}_{-1.719}$ & $13.021^{+0.265}_{-0.376}$ & $13.143 \pm 0.139$ & $-0.121$ & $13.102^{+0.174}_{-0.085}$ & $-0.080$\\
4.1795 & $7.413^{+1.804}_{-1.804}$ & $12.551^{+0.269}_{-0.374}$ & $12.739 \pm 0.126$ & $-0.188$ & $12.708^{+0.156}_{-0.087}$ & $-0.157$\\
5.2888 & $10.755^{+2.265}_{-2.265}$ & $12.506^{+0.232}_{-0.306}$ & $12.380 \pm 0.128$ & $0.126$ & $12.340^{+0.168}_{-0.071}$ & $0.166$\\
\hline

\multicolumn{7}{c}{MMH $= 100$, Field $= 5000$}\\
\hline
0.0000 & $2.662^{+1.057}_{-1.057}$ & $14.261^{+0.337}_{-0.655}$ & $14.397 \pm 0.185$ & $-0.137$ & $14.347^{+0.226}_{-0.123}$ & $-0.086$\\
1.0779 & $6.118^{+2.501}_{-2.501}$ & $14.264^{+0.306}_{-0.541}$ & $13.973 \pm 0.159$ & $0.291$ & $13.929^{+0.178}_{-0.100}$ & $0.334$\\
2.0700 & $7.027^{+2.809}_{-2.809}$ & $13.701^{+0.332}_{-0.578}$ & $13.538 \pm 0.147$ & $0.163$ & $13.492^{+0.184}_{-0.087}$ & $0.209$\\
3.0604 & $7.524^{+3.073}_{-3.073}$ & $13.170^{+0.372}_{-0.655}$ & $13.143 \pm 0.139$ & $0.028$ & $13.102^{+0.174}_{-0.085}$ & $0.069$\\
4.1795 & $8.909^{+4.015}_{-4.015}$ & $12.779^{+0.426}_{-0.799}$ & $12.739 \pm 0.126$ & $0.040$ & $12.708^{+0.156}_{-0.087}$ & $0.071$\\
5.2888 & $9.164^{+3.408}_{-3.408}$ & $12.302^{+0.395}_{-0.658}$ & $12.380 \pm 0.128$ & $-0.078$ & $12.340^{+0.168}_{-0.071}$ & $-0.038$\\
\hline

\multicolumn{7}{c}{MMH $= 50$, Field $= 5000$}\\
\hline
0.0000 & $2.668^{+1.825}_{-1.825}$ & $14.263^{+0.505}_{-2.400}$ & $14.397 \pm 0.185$ & $-0.134$ & $14.347^{+0.226}_{-0.123}$ & $-0.084$\\
1.0779 & $6.034^{+3.349}_{-3.349}$ & $14.251^{+0.390}_{-0.892}$ & $13.973 \pm 0.159$ & $0.278$ & $13.929^{+0.178}_{-0.100}$ & $0.321$\\
2.0700 & $6.825^{+3.559}_{-3.559}$ & $13.671^{+0.413}_{-0.883}$ & $13.538 \pm 0.147$ & $0.133$ & $13.492^{+0.184}_{-0.087}$ & $0.179$\\
3.0604 & $6.779^{+4.954}_{-4.954}$ & $13.049^{+0.597}_{-2.079}$ & $13.143 \pm 0.139$ & $-0.093$ & $13.102^{+0.174}_{-0.085}$ & $-0.052$\\
4.1795 & $8.217^{+5.625}_{-5.625}$ & $12.681^{+0.599}_{-1.792}$ & $12.739 \pm 0.126$ & $-0.058$ & $12.708^{+0.156}_{-0.087}$ & $-0.028$\\
5.2888 & $8.242^{+5.182}_{-5.182}$ & $12.160^{+0.614}_{-1.627}$ & $12.380 \pm 0.128$ & $-0.220$ & $12.340^{+0.168}_{-0.071}$ & $-0.180$\\
\hline

\multicolumn{7}{c}{MMH $= 25$, Field $= 5000$}\\
\hline
0.0000 & $2.618^{+2.095}_{-2.095}$ & $14.242^{+0.567}_{\rm unconstrained}$ & $14.397 \pm 0.185$ & $-0.155$ & $14.347^{+0.226}_{-0.123}$ & $-0.104$\\
1.0779 & $6.122^{+4.839}_{-4.839}$ & $14.264^{+0.503}_{-2.175}$ & $13.973 \pm 0.159$ & $0.291$ & $13.929^{+0.178}_{-0.100}$ & $0.335$\\
2.0700 & $8.220^{+4.613}_{-4.613}$ & $13.860^{+0.415}_{-0.938}$ & $13.538 \pm 0.147$ & $0.322$ & $13.492^{+0.184}_{-0.087}$ & $0.368$\\
3.0604 & $7.442^{+5.644}_{-5.644}$ & $13.158^{+0.597}_{-2.221}$ & $13.143 \pm 0.139$ & $0.015$ & $13.102^{+0.174}_{-0.085}$ & $0.056$\\
4.1795 & $9.223^{+6.915}_{-6.915}$ & $12.821^{+0.619}_{-2.175}$ & $12.739 \pm 0.126$ & $0.082$ & $12.708^{+0.156}_{-0.087}$ & $0.113$\\
5.2888 & $9.357^{+7.165}_{-7.165}$ & $12.329^{+0.683}_{-2.545}$ & $12.380 \pm 0.128$ & $-0.051$ & $12.340^{+0.168}_{-0.071}$ & $-0.011$\\
\hline
\end{longtable}

% Restore defaults
\renewcommand{\arraystretch}{1.0}
\setlength{\tabcolsep}{6pt}

% Return to two-column text only after BOTH tables
\twocolumn
%-------------------------------------------------
%-------------------------------------------------

%-------------------------------------------------
%-------------------------------------------------

25 to 50 objects reduces the mass uncertainty by a third, and a further doubling by a similar amount. Effort spent enlarging the smallest samples is repaid most strongly, but the gains become substantially smaller at larger sample sizes.

The step from 500 to 1000 is not a like-for-like doubling: samples of 500 and below are averaged over 100 random draws, while those of 1000 and above are single deterministic selections (Section~\ref{sec:halo_samples}), so the gain measured across that step mixes sample size with estimator. The trend on either side does not, and the contrast between the two regimes stands.

These are not academic sample sizes. Earlier work on $z \sim 4$ quasars showed that small samples can still place useful constraints on host-halo mass through clustering, while also exposing the limitations imposed by sample size and the quasar--halo connection \citep{White2008}. \citet{Arita2023} derive a halo mass for quasars at $z \approx 6$ from a clustering analysis of 107 objects, obtaining a bias of $20.8 \pm 8.7$ and $\log (M/h^{-1}\mathrm{M}_{\odot}) = 12.70^{+0.39}_{-0.70}$. Our 100-object configurations return $\pm 0.31$ dex from the clustering measurement alone, of comparable magnitude, though the two differ in survey volume, selection, reference population and estimator. The comparison is between measurement errors only. An observer can quantify the noise in a clustering measurement but not the offset between the mass it returns and the true mean mass of the population, measured here at a median of $0.103$ dex, with a further $0.07$ dex from the calibration of the bias--mass relation (Section~5.1). Neither term is removed simply by increasing the sample size, and neither is captured by the clustering measurement error, so reading that uncertainty as the full error budget would understate the total uncertainty on the recovered mass. At higher redshift the samples are smaller still: quasar--galaxy clustering has been constrained at $z \simeq 7.3$ from eight companion galaxies across two quasar fields \citep{Schindler2026}. For a survey holding a few tens of objects, modest additional observation can therefore produce substantial gains; those gains diminish considerably once samples reach several hundred.

The $J = 100$ samples highlight a second issue. A draw of 500, 100, 50 or 25 from a parent population of 1000 is a survey that has recovered half, a tenth, a twentieth or a fortieth of the rare objects present. Read this way the uncertainties describe the cost of random incompleteness rather than simply of small numbers: a survey recovering half its target population loses little, while one recovering a twentieth loses a great deal.

Two qualifications limit how directly these figures transfer. The draws considered here are random, and represent an idealised case. Selection correlated with halo or galaxy properties, as any flux- or mass-limited selection will be, can alter both the precision of the measurement and the population whose characteristic mass is inferred. The averaging over 100 draws is also an ensemble operation, available to a simulation but not to an observer holding a single sample: it characterises the expected behaviour of a sample of that size rather than the uncertainty on one realisation of it.

%-------------------------------------------------
%-------------------------------------------------

\subsection{Reference-sample variance}
\label{sec:reference_variance}

The field sample against which clustering is measured is a distinct source of error in our analysis, and it behaves differently from the massive sample. An error in the massive sample affects only the measurement made with it. An error in the field enters every cross-correlation measured against that field, in the same direction and by a similar amount, and observing more rare objects does not reduce it.

Section~\ref{sec:field_reference} shows this directly, in one low realisation of the 5000-halo field at $z=1.08$. The direction of the resulting displacement is not accidental. From Equations~(\ref{eq:bias_acf}) and (\ref{eq:bias_massive}) the massive-halo bias varies inversely with the square root of the field auto-correlation, a convex function, so symmetric noise in the field measurement inflates the expected bias: for the 5000-halo field, with a 31 per cent mean uncertainty on $\xi(8)$, the leading-order inflation is 3.6 per cent in bias, or $0.04$ dex in mass, if the field fluctuation is treated as independent of the cross-correlation. Consistent with this, the 5000-halo configurations return a mean offset of $+0.048$ dex against $-0.065$ to $-0.084$ dex for the larger fields, though roughly half the excess above the estimate comes from the single low realisation at $z = 1.08$. One field realisation per size and redshift cannot establish this as a systematic, but the mechanism predicts that a small reference sample will tend to bias the recovered mass high rather than merely scatter it.

Cross-correlation against a denser tracer is already the strategy of choice for rare populations at high redshift, since it retains signal where the rare population's own auto-correlation does not. Finite survey volumes are known to introduce cosmic variance into measurements of high-redshift populations \citep{Trenti2008,Moster2011}. The effect isolated here is distinct: uncertainty in the measured clustering amplitude of the reference sample propagates coherently into every rare-population mass inferred from that reference. This is relevant to current measurements, where reference populations can be small---\citet{Schindler2026}, for example, constrain the galaxy reference clustering from 51 galaxies across two quasar fields.

Rare massive objects are expensive to observe, while suitable reference populations are generally easier to obtain: they can often be drawn from the same imaging that identified the rare objects, or from existing wide surveys of the same area. The reference nonetheless carries an error that cannot be reduced from the massive side. Increasing the rare-object sample alone is therefore insufficient once uncertainty in the reference clustering becomes appreciable; the reference population must itself be measured with adequate precision.

Our data and results cannot locate a threshold below which reference-sample variance becomes limiting. Four field sizes at six redshifts establish that the effect exists and that it is independent of the rare-object sample.

%-------------------------------------------------
%-------------------------------------------------

\subsection{A halo-only benchmark}
\label{sec:halo_only}

Working directly with host-halo masses removes the additional scatter introduced by mapping observed galaxy properties onto halo mass. The offsets measured here are those of the clustering method alone, applied to a population whose true masses are known exactly.

The magnitudes make that isolation worth having. Intrinsic scatter in stellar mass at fixed halo mass is typically $\lesssim 0.2$~dex for massive systems \citep{2018ARA&A..56..435W}, while the method's own offsets are of order $0.10$~dex. In a galaxy sample the two contributions would be comparable, and could not be separated from the observations alone.

For galaxies the inferred mass must contend with that scatter in addition to the measurement error quantified here. Since the recovery of halo mass improves towards higher redshift in the halo-only case (Section~\ref{sec:recovery_accuracy}), any deterioration observed for a galaxy population at early times would point to additional effects associated with the tracer--halo connection or with observational selection, rather than to the halo-clustering method alone. That case will be examined in a forthcoming paper.

Our method makes no reference to what the rare objects are. It requires positions and a reference population, and can in principle be applied to massive quiescent galaxies, submillimetre galaxies, luminous quasars, and the compact red sources now being identified at high redshift by JWST, whose clustering is beginning to be measured \citep{Carranza25}. The accuracy achieved for any such population will depend on its selection and on how it occupies haloes, which is what the halo-only case sets aside.

%-------------------------------------------------
%-------------------------------------------------

\subsection{Limitations}
\label{sec:limitations}

The Millennium simulation adopts WMAP1 parameters \citep{Spergel2003,Springel2005}, with $\sigma_8 = 0.9$ against the $0.81$ preferred by Planck \citep{2020A&A...641A...6P}. Bias at fixed halo mass depends on $\sigma_8$, so the values quoted here are specific to this cosmology. They should be read as an error budget for the method rather than as a correction table for application to observations.

The \citet{Tinker2010} relation carries uncertainties beyond the calibration scatter discussed in Section~\ref{sec:sample_size}. It is an empirical fit to halo bias measured in simulations, and departures from a universal bias relation remain, particularly for rare haloes. Our samples occupy precisely this high-mass regime, so uncertainties in the adopted bias--mass calibration propagate directly into the inferred masses. A further limitation is assembly bias: haloes of the same mass can cluster differently depending on their assembly history, an effect demonstrated in the Millennium simulation itself \citep{Gao2005, Croton2007}.

The redshift dependence of the offset carries a similar qualification. The change from under-recovery at low redshift to over-recovery at high redshift follows from the shape of the bias--mass relation over the mass range occupied by these samples, so the sign of the offset depends on where the population lies on that relation. In our Millennium setup the offset is negative at every redshift up to $z = 4.18$ and positive at $z = 5.29$. With one cosmology and the limited time resolution of the Millennium snapshots, we cannot determine exactly where the sign change occurs or whether its location is general.

The uncertainties come from a 27-region jackknife within a single $500\,h^{-1}\,\mathrm{Mpc}$ box. \citet{2009MNRAS.396...19N} find that internal estimators of this kind do not reproduce external error estimates across the range $1$--$25\,h^{-1}\,\mathrm{Mpc}$. Our measurements at $r = 8\,h^{-1}\,\mathrm{Mpc}$ fall within the range over which internal and external estimates need not agree, so the quoted jackknife uncertainties should themselves be regarded as approximate. They also measure the variance of the estimator within one volume rather than the cosmic variance between independent volumes, which for a survey of comparable size would add a further contribution \citep{Trenti2008, Moster2011}.

Finally, the measurements use the three-dimensional positions available in a simulation. Real surveys work in projection, with angular positions and redshifts of finite precision, both of which alter the clustering measurement. The accuracies established here are those achievable in the absence of these effects, and quantifying the impact of projection and photometric redshift uncertainty is required before they can be applied to a real survey.

\section{Summary and Future work}
\label{sec:summary}

We have measured how accurately the mean logarithmic halo mass of a rare, massive population can be recovered from clustering alone, using dark-matter haloes in the Millennium simulation, where the true masses are known. Massive samples of 25 to 10000 haloes were cross-correlated against field references of 5000 to 50000 haloes at six redshifts between $z=0$ and $z=5.29$, the bias measured at $r=8\,h^{-1}\,\mathrm{Mpc}$, and the \citet{Tinker2010} relation inverted to return a characteristic mass. Across 192 configurations, comparing the recovered mass with the mean logarithmic halo mass measured directly from the simulation gives the following:

\begin{itemize}

\item The characteristic mass is recovered to a median of $0.103$ dex, with roughly two thirds of configurations falling within $0.14$ dex.

\item Accuracy and precision do not improve together. The offset depends only weakly on the number of massive objects, while the uncertainty grows from $\pm0.08$ dex at 10000 haloes to $\pm0.71$ dex at 25. Larger samples are measured more precisely but extend further down the mass function, broadening the population that a single characteristic mass represents.

\item The return on enlarging a sample falls steeply with its size. Doubling a sample from 25 to 50 objects reduces the mass uncertainty by about a third, while doubling from 5000 to 10000 reduces it by only about a tenth. Equivalently, recovering half of a target population loses little, while recovering a twentieth loses a great deal.

\item The field reference sample is a distinct source of error. Its uncertainty enters every cross-correlation measured against it coherently and cannot be reduced by observing more rare objects. A single low realisation of the smallest field displaced every massive sample paired with it, from 25 haloes to 10000, in the same direction and by nearly the same amount.

\item Recovery improves towards higher redshift, with the offset changing from under-recovery at low redshift to over-recovery above $z \approx 5$, following the changing shape of the bias--mass relation across the mass range occupied by the samples.

\item The calibration of the bias--mass relation contributes a further systematic of ${\sim}0.07$ dex, comparable to the measurement uncertainty of the best configurations. Statistical precision beyond this point increasingly exposes the calibration uncertainty rather than improving the inferred mass.

\end{itemize}

Taken together, these results carry a practical message for survey design. Enlarging a sample of a few tens of rare objects is well repaid; enlarging one of several hundred is not, and beyond that point the reference population and the bias--mass calibration set what can be achieved.

Several avenues follow from this work.

\begin{itemize}

\item Applying the same measurement to galaxies will quantify the additional error introduced by the galaxy--halo connection. Because the halo-only recovery improves towards higher redshift, any deterioration for a galaxy population at early times would point to the tracer--halo connection or to selection rather than to the clustering method itself. That analysis is in preparation.

\item The measurements here use the three-dimensional positions available in a simulation, while a real survey works in projection with redshifts of finite precision. Projection and redshift uncertainty dilute the measured clustering amplitude and make the recovered mass depend on how well the redshift distribution is known. Quantifying that degradation is required before the accuracies established here can be applied to a real survey. That work is under way.

\end{itemize}

Observationally the method requires only positions and a reference population, and applies to any rare massive population for which both are available: massive quiescent galaxies, submillimetre galaxies, luminous quasars, and little red dots now being identified at high redshift. For such populations the results presented here establish a halo-only benchmark for what the clustering method can deliver, before the connection between those objects and their haloes is considered.

\section*{Acknowledgements}
This research was supported in part by the Australian Government through the Australian Research Council (ARC) Centre of Excellence for Dark Matter Particle Physics (CDM, CE200100008). AB would like to thank Prof. Karl Glazebrook for many helpful discussions throughout the development of this project. AB also thanks Atrideb Chatterjee and Manodeep Sinha for valuable discussions. DC acknowledges the support of an ARC Future Fellowship (FT220100841). This work was performed on the Ngarrgu Tindebeek/OzSTAR national facility at Swinburne University of Technology. The OzSTAR program receives funding in part from the Astronomy National Collaborative Research Infrastructure Strategy (NCRIS) allocation provided by the Australian Government, and from the Victorian Higher Education State Investment Fund (VH-ESIF) provided by the Victorian Government.

This work made use of the \textsc{SciPy} \citep{Virtanen2020},
\textsc{NumPy} \citep{Harris2020}, and
\textsc{Matplotlib} \citep{Hunter2007} packages for Python.

%%%%%%%%%%%%%%%%%%%%%%%%%%%%%%%%%%%%%%%%%%%%%%%%%%
\section*{Data Availability}

The Millennium data are publicly accessible through the Millennium database. The data products of this work can be shared upon request to the corresponding author.

% \newpage

\bibliographystyle{mnras}
\bibliography{reference} % if your bibtex file is called example.bib

%\begin{thebibliography}{99}
%\end{thebibliography}

%%%%%%%%%%%%%%%%%%%%%%%%%%%%%%%%%%%%%%%%%%%%%%%%%%

%%%%%%%%%%%%%%%%% APPENDICES %%%%%%%%%%%%%%%%%%%%%

% \appendix

% \section{Some extra material}

% If you want to present additional material which would interrupt the flow of the main paper,
% it can be placed in an Appendix which appears after the list of references.

%%%%%%%%%%%%%%%%%%%%%%%%%%%%%%%%%%%%%%%%%%%%%%%%%%

% Don't change these lines
\bsp	% typesetting comment
\label{lastpage}
\end{document}